\documentclass[5p]{elsarticle} % Aptara syntax

\usepackage[ruled]{algorithm2e}
\usepackage{multirow}
\usepackage{pbox}
\usepackage{color}
\usepackage{subfig}
\usepackage{mathtools}
\usepackage{tablefootnote}

\DeclareGraphicsExtensions{.pdf,.png,.jpg,.eps}
\graphicspath{{./}{./figs/}}
\usepackage[]{changes}
\definechangesauthor[color=red]{AP}
\definechangesauthor[color=blue]{VA}
\usepackage{soul}
\sethlcolor{yellow}

\usepackage[]{todonotes}
\begin{document}

\title{Ego Network Structure in Online Social Networks and its Impact on
Information Diffusion}
\author[iit]{Valerio Arnaboldi\corref{cor1}} \ead{v.arnaboldi@iit.cnr.it}
\author[iit]{Marco Conti} \ead{m.conti@iit.cnr.it}
\author[iit]{Massimiliano La Gala} \ead{m.lagala@iit.cnr.it}
\author[iit]{Andrea Passarella} \ead{a.passarella@iit.cnr.it}
\author[iit]{Fabio Pezzoni} \ead{f.pezzoni@iit.cnr.it}
\cortext[cor1]{Corresponding author}
\address[iit]{IIT-CNR, Via G. Moruzzi 1, 56124, Pisa, Italy}

\begin{abstract}

In the last few years, Online Social Networks (OSNs) attracted the interest of a
large number of researchers, thanks to their central role in the society.
Through the analysis of OSNs, many social phenomena have been studied, such as
the viral diffusion of information amongst people. What is still unclear is the
relation between micro-level structural properties of OSNs (i.e. the properties
of the personal networks of the users, also known as ego networks) and the
emergence of such phenomena. A better knowledge of this relation could be
essential for the creation of services for the Future Internet, such as highly
personalised advertisements fitted on users' needs and
characteristics. In this paper, we contribute to bridge this gap by analysing
the ego networks of a large sample of Facebook and Twitter users. Our results
indicate that micro-level structural properties of OSNs are interestingly
similar to those found in social networks formed offline. In particular, online
ego networks show the same  structure found offline, with social
contacts arranged in layers with compatible size and composition. From the
analysis of Twitter ego networks, we have been able to find a direct impact of
tie strength and ego network circles on the diffusion of information in the
network. Specifically, there is a high correlation between the frequency of
direct contact between users and her friends in Twitter (a proxy for tie
strength), and the frequency of retweets made by the users from tweets generated
by their friends. We analysed the correlation for each ego network layer
identified in Twitter, discovering their role in the diffusion of
information.

\end{abstract}


\maketitle

\section{Introduction}
\label{sec:intro}

The impressive penetration of Internet technologies and the establishment of
participatory forms of content generation and exchange, such as the Web 2.0
paradigm, paved the way for the diffusion of Online Social Networks (OSNs). OSNs
are nowadays a significant part of the prosumer paradigm shift in
communication and data exchange, whereby users can actively create and share
information with each other, rather than being passive consumers of contents as
in more traditional media. In addition, OSNs are becoming one of the preferred
ways to manage social relationships for an increasing number of people and their
use is growing also far beyond supporting social relationships between people.
Today OSNs are already successfully used, among
others, for commercial
recommendations, online content curation, advertising, and much more.

OSNs are significantly contributing to the so called cyber-physical world (CPW)
convergence~\cite{conti2012looking}, which envisions a world where people
actions and interactions in the cyber (virtual) world, enabled by ICT, and in the
physical world are strongly dependent upon each other and knitted into a single
whole. In the CPW, actions taken in the virtual world are directly transferred
to the physical world and vice versa. For example, social relationships in OSNs
(i.e. in the virtual world) often depend upon those existing in the physical
world and actions taken in OSNs modify the state of the physical world
(e.g., mass movements or rallies that are organised and advertised
exclusively over OSNs).

Characterising the properties of OSNs has been a very active research topic
recently. The scale of these networks make this task challenging per se, and the
diffusion of OSN services makes results obtained from this type of analysis
impactful. In addition, the study of OSN structural properties is fundamental
for the creation of a series of new services for the Future Internet highly
customised on the user's characteristics and needs. For example,
the structure
of OSNs can be exploited to develop new efficient and cost-effective marketing strategies, as
shown in~\cite{Domingos2001Mining}. Despite this, most of the analyses
conducted so far on OSNs are focused on macro-level structural properties only
(e.g. global clustering coefficient, diameter, presence of communities), whereas
micro-level structures (i.e. the properties of personal social networks of the
users) have not been investigated in detail. In sociology and anthropology, the
micro-level structures of social networks formed offline (not mediated by the
use of the Internet) are found to be directly related to most of the social
phenomena arising in the network and, therefore, it can be reasonably expected
that micro-level structures in OSNs could impact on the aforementioned OSN
services.

The chief aim of this paper is twofold. On the one hand, we investigate in
detail key properties of OSNs at the micro level. On the other hand, we show how
these properties determine patterns of information diffusion in OSNs. In this
way, we contribute to discover the relation between microscopic and macroscopic
properties of OSNs, where the former are related to the social behaviour and
structure of individual users, while the latter
  relate to social phenomena
involving the network as a whole.

In order to characterise the micro-level structures of OSNs, we focused our
analysis on ego networks. An ego network is defined as a portion of a social
network formed of a given individual, termed \emph{ego}, and the other persons
with whom she has a social relationship, termed \emph{alters}. Ego
  networks have been the subject of a very significant body of work in the
  sociology and anthropology literature, that has characterised some of their
  fundamental properties (see Section~\ref{sec:microlevel} for more details). One of the most important is the presence, in
the ego network structure, of a series of concentric layers of alters with
different levels of intimacy and size. The key parameter to distinguish
  between alters at different layers is the \emph{tie strength} of the social
  relationship with the ego, which is typically approximated with the frequency
of contact between them (a more precise description of these
results is presented as background material in Section~\ref{sec:background}).

Section~\ref{sec:analysis} of this paper reports an analysis that investigates
whether similar ego network structures can also be found in OSNs. This analysis allowed us
to assess the differences between the baseline results in social sciences and
the properties we have observed in different OSN data sets (described in
Section~\ref{sec:datasets}) obtained from Facebook and Twitter. Notably, we have
found a layered structure in OSN ego networks similar to the one identified in
offline social networks in terms of: (i) number of layers, (ii) frequency of
contact of the layers, and (iii) scaling factor between the size of adjacent
layers. This indicates that, as far as the structural properties of social
relationships are concerned, human social behaviour seems to be unaltered by the
use of OSNs. This further confirms the existence of the CPW convergence and it
must be taken into account for the creation of user-centric
future-Internet services.

Starting from these results, in Section~\ref{sec:diffusion_analysis},
we report an analysis aimed at
assessing the role of the ego network structure on the diffusion of information.
We analysed the impact of tie strength and the presence of ego network circles
on information propagation in Twitter ego networks, for which we have complete
information on the creation of tweets and retweets. In accordance with the
literature (see for example~\cite{Bakshy2012}), we found that
\emph{weak}
  ties, associated with lower levels of direct interactions than
  \emph{strong} ties, also
transport a lower number of retweets. Despite this, the high number of weak ties
in the ego networks makes the total amount of information circulating through them
exceed the amount of information passing through strong ties.
Then, we analysed the correlation between the frequency
of interaction between users in Twitter (a proxy for their tie strength) and the
frequency of retweets that flow
  through the social links that connect these users. The correlation has a medium/high
  value ($r = 0.46$), but it is not sufficient to justify a model able to predict
  information diffusion from tie strength. Hence, we further investigated this aspect on two axes: (i) by studying the
  correlations within single ego network layers, and (ii) by dividing social
  relationships in two classes, the first related to alters who use Twitter for
  socialising and the second containing other types of alters, like companies, public
  figures, etc. The results indicate that the correlation between tie strength and
  information diffusion increases when we move from the outer to the inner parts of ego
  networks (from weak to strong ties), with values greater than $0.6$ for the innermost
  layer. Perhaps more surprisingly, the correlations for both classes of
  alters are
  sensibly higher than those found when the two are mixed together, with the first
  class (i.e. people with social behaviour) showing the higher values of correlation
  (close to $0.8$ for the innermost layer and always higher than $0.6$ for the other
  layers). 

Therefore, in summary, the results presented in the paper show a significant
similarity between social network structures in online and offline environments.
Not only the structure of ego networks is remarkably similar, but also these
structures significantly impact on the way OSN services are used. Specifically,
we have found that the patterns
of information diffusion can be explained quite precisely starting from ego
network structures of the individual users.

\section{Background and Motivations}
\label{sec:background}

Social networks are
structures composed of a set of social actors (e.g.  individuals, organisations)
and a set of ties (i.e. social relationships) connecting pairs of these actors.
They are usually expressed in the form of graphs consisting of nodes
representing social actors connected by edges, or arcs, which represent social
relationships. We define \emph{online} social networks as the social networks in
which social relationships are maintained by the use of the Internet (e.g.
Facebook, Twitter, e-mail exchange networks), and \emph{offline} social networks
as social networks formed outside the Internet, for example, face-to-face
communication networks or phone call networks.

Both offline and online social networks have been analysed as typical
  complex network systems~\cite{newman2003structure}, i.e. with the same
  methodology used for other kinds of networks, such as biological and
  technological networks. Indeed, they have shown to present well-known
  properties of complex networks, such as the \emph{small-world
  property}~\cite{Travers1969}. As discussed in~\cite{Watts1998collective}, a
  small-world network is characterised by short average distances between any
  two nodes connected via a chain of intermediate links. In addition, a small world network shows a high level of clusterisation
  (or network transitivity) compared to a random network, where
  clusterisation is the probability that two neighbours connected to a
  node will also be connected to each other. The small-world property directly
  impacts on the ability of the network to spread information quickly.
  Not surprisingly, another typical property found in offline and online
    social networks is the presence of communities~\cite{Newman2003}. These
    studies are normally carried out considering the unweighted network graph in
    which each edge (or arc) represents the mere existence of a social
    relationship without including information that can distinguish between
    different types of relationships. This is due to the fact that information
    about social relationships is not trivial to infer since it normally refers
    to qualitative aspects that are difficult to measure. Nevertheless, in
    particular for the analysis of microscopic social network properties,
    characterising (and distinguishing between) different types of social links
    is fundamental. In particular, \emph{tie strength}, i.e. a quantitative
  measure of the importance of the social links linking two people, is a very
important parameter, which has been widely investigated in the sociology and
anthropology literature.

\subsection{Measures of Tie Strength in Social Networks}
\label{sec:tsosn}

In his seminal work, Mark Granovetter informally defined the strength of a
social relationship as a linear combination of time, emotional intensity,
intimacy, and reciprocal services~\cite{Granovetter2007}. Social relationships
can be roughly divided into strong and weak ties, where the former denote more
important relationships and the latter represent acquaintances. Besides their
lower strength, weak ties are generally more numerous than
strong ties. For this reason, the cumulative strength of weak ties could exceed
that of strong ties and their impact on social phenomena could be substantial.
Other measures of tie strength have been successively constructed and validated
by Peter Marsden in~\cite{Marsden2011}.

Based on the results found by Marsden, several techniques to measure tie
strength have been proposed also for OSNs, for example
in~\cite{Gilbert2009,Gilbert2012,Arnaboldi2013a,Kahanda2009}. These studies
indicate that tie strength can be effectively estimated using some measurable
indicators. In particular, the frequency of contact seems to be the best
among them, especially in online environments, also
considering that it is easy to obtain from online communications logs. This
has been confirmed in a study on Facebook~\cite{Jones2012Inferring}, where the
authors asked a set of Facebook users to name their closest friends in real
life, and they found that contact frequency can be used to accurately
discriminate closest friends from acquaintances.

\subsection{OSN Analysis based on Tie Strength}

Recently, some work has been done to characterise the differences between strong
and weak ties in OSNs, and to relate them with observable properties of the
networks. Ties connecting otherwise disconnected parts of the network (also
known as bridges) are associated to lower interaction levels than ties
connecting clusterised parts of the network~\cite{Volkovich2012TheLength}. This
has been observed also in phone call networks~\cite{Onnela2007}, and is
consistent with the \emph{strength of weak ties} hypothesis of
Granovetter~\cite{Granovetter2007} that postulates that social links
connecting distant and otherwise disconnected parts of a social network must be
weak ties. In~\cite{gong2012evolution,Wilson2012} two studies on Facebook and
Google+ show that considering tie strength in the analysis of social network structures
reveals properties that are not visible from the unweighted networks, and can
lead to significantly different results. For example, the average
  distance between nodes in the Facebook network, that is less than $4$ in the unweighted
  network, is between $5$ and $10$ when tie strength is considered. This is because
  many links in OSNs appear to be inactive, and considering them in the structural analysis of
  OSNs can lead to inconsistent or wrong results. For example, if we consider information
  diffusion, no information passes through inactive links, and they cannot be considered
effective channels for information diffusion.

In addition, a relation between geographical location and tie strength
has been found, with strong ties connecting most of the time people in physical
proximity, and bridges connecting part of the network far from each
other~\cite{Volkovich2012TheLength,Leskovec2007}. Mobility also plays an
important role in the formation of social ties and in determining their
strength, since meeting other people enables social
interactions~\cite{Grabowicz2014Entangling}.

\subsection{Ego Network Model}
\label{sec:microlevel}

\begin{figure}[t]
  \centering
  \includegraphics[width=0.4\textwidth]{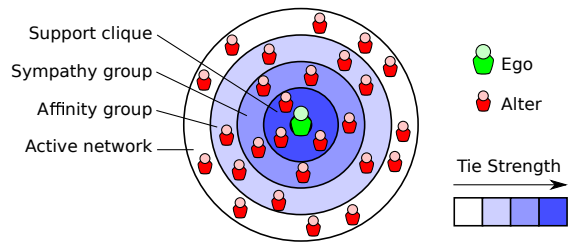}
  \caption{Ego network structure.}
  \label{fig:egonet}
\end{figure}

In order to study the micro-level structural properties of social networks,
researchers defined the \emph{ego network}, as a simple social network model
formed of an individual (called \emph{ego}) and all the persons with whom the
ego has a social link (\emph{alters}). In an ego network, alters are normally
arranged in a series of four or five inclusive groups
(called \emph{circles}) according to the strength of their social ties.
Figuratively, an individual ego can be envisaged as sitting at the centre of the
series of concentric circles~\cite{Roberts2010} as depicted in
Figure~\ref{fig:egonet}.  Each of these circles has typical size and tie
strength. The latter is usually estimated using the frequency of contact between
the ego and the alters.  Note that in the following,
  as typically done in the literature, a \emph{circle} also contains all alters of the more internal
  circles (with higher tie strength), while a \emph{ring} only contains alters
of a given circle that are not part of any more internal circle.

The first circle, called \emph{support clique}, contains alters with very strong
social relationships with the ego, informally identified in the literature as
\emph{best friends}. These alters are people contacted by the ego in case of a
strong emotional distress or financial disasters. The size of this circle is
limited, on average, to $5$ members, usually contacted by the ego at least once
a week. The second circle, called \emph{sympathy group}, contains alters who can
be broadly identified as \emph{close friends}. This circle is formed of, on
average, $15$ members contacted by the ego at least once a month. The next
circle is the \emph{affinity group} (or \emph{band} in the ethnographic
literature), which contains about $50$ alters usually representing causal
friends or extended family members~\cite{Roberts2009}. Although some studies
tried to identify the typical frequency of contact of this circle, there are no
accurate results in the literature about its properties, due to the
difficulties related to the manual collection of data about the alters contained
in it through interviews or surveys. The last circle in the ego network model is
the \emph{active network}, which includes all the other circles, for a total of
about $150$ members. This circle contains people for whom the ego actively
invests a non-negligible amount of resources to maintain the related social
relationships over time. People in the active network are contacted, by
definition, at least once a year. The active network size coincides with the
\emph{Dunbar's number}, that identifies the average limit of the number of
social relationships an individual can actively maintain due to cognitive
constraints of the brain and the limited time for
socialising~\cite{Dunbar1998hyp}. Alters beyond the active network are
considered inactive, since they are not contacted regularly by the ego. These
alters are grouped in additional external circles called \emph{mega-bands} and
\emph{large tribes}. One of the most stunning facts about ego network circular
structure is that the ratio between the size of adjacent circles appears to be a
constant with a value around $3$, and this holds true for ego
  networks of users belonging to various social environments, as shown in~\cite{Zhou2005}.
For a complete discussion about the properties of the ego network circles we
refer the reader to~\cite{Sutcliffe2012}.

\subsection{Analyses on Ego Networks in OSNs}

As far as the structure of ego networks in OSNs are concerned, and in particular
the ego network model applied to online environments, still little is known. In
the following, we summarise the preliminary results found on ego network
properties of OSNs, which represent the starting point of this work.

The authors of~\cite{Arnaboldi2013a} and~\cite{Goncalves2011pone} analysed
two data sets from Facebook and Twitter respectively, founding evidences of the
presence of the Dunbar's number in OSNs. Its presence indicates that, even
though Facebook and Twitter allow people to have thousands of online social
contacts, they only maintain a limited set of active relationships.

The authors of~\cite{Arnaboldi2013cosn}
analysed a large data set of Twitter
communications discovering that the limited capacity people have for socialising
bounds the amount of contacts they can actively maintain over time, as defined
in the ego network model.

In a recent analysis on Twitter communication data, it has been found that the
structure of Twitter ego networks is directly related to the network status of
egos (defined as the ratio between followers and following), to the topic
diversity of the tweets generated by egos, and to geography~\cite{Quercia2012b}.
In particular, egos who have contacts spanning structural holes (gaps between
separated groups of people) have higher network status, higher topic diversity,
and more geographically sparse networks than egos with highly clusterised
networks. This validates the idea that social capital is created by bridging
structural holes, as proposed by sociologist Ron Burt~\cite{Burt2001}. This has been
further confirmed by the work presented in~\cite{Aral2007TheDiversity}, where
the authors identify a trade-off between the diversity of information that a
network can provide and the average strength of its social ties (called
bandwidth). This trade-off is connected to the ability of egos to bridge
structural holes. In a highly clusterised network the diversity of acquirable
information is low, but bandwidth is high, with a positive impact on the
quantity and quality of resources that ego can obtain from her intimate alters.
On the other hand, in ego networks where egos bridge many structural holes, the
diversity of information is higher, to the detriment of network
bandwidth. This also confirms results in the sociology literature, which
  show that the higher the number of social links (and, therefore, the higher
  the accessible information diversity), the lower the average strength of
  social ties. This is, again, related to the fact that humans can only allocate
  a finite amount of cognitive resources for socialising, and therefore a higher
  number of social relationships translates into a lower average tie strength.
  Note that this very same phenomenon has also been observed in
  Twitter~\cite{Goncalves2011pone}.

Although these results give a first insight on the properties of ego networks in
OSNs, there is still a lack of knowledge about the micro-level structural
properties of OSNs. Specifically, it is not clear if structures similar to those
described by the ego network model could be found also in OSNs (e.g. the
presence of ego network circles and their properties). In this paper, we aim to
bridge this gap by providing a solid analysis of the ego network structures in
OSNs. 

This paper extends our initial analyses on Facebook and
  Twitter~\cite{Arnaboldi2012,Arnaboldi2013,Dunbar2015TheStructure}. Specifically, in this
  paper we provide a more detailed analysis of ego network structures in both
  Online Social Networks, and we exploit it to characterise the impact of these
structures on information diffusion.

\subsection{Analyses on Information Diffusion in OSNs}

In this paper, we also investigate the role of
the structural properties of OSN ego networks in the diffusion of information,
one of the most important macro-level phenomena in social networks.
Specifically, we study information diffusion in Twitter ego networks and we
investigate the role of tie strength and the ego network structure in the
process.

Prior work on information diffusion has focused on detecting collective
behaviour in the network, such as the formation of information cascades (i.e.
the epidemic diffusion of information triggered by the observation and the
adoption of the behaviour of others). Some works successfully predicted the
outbreak of cascades in social
networks~\cite{Cui2014Cascading,Leskovec2007Cost}. Other papers show that
different types of users have different roles in diffusing
information (e.g. normal users, opinion leaders, mass media
sources)~\cite{Cha2012TheWorld}. Mass media diffuse most of the information, but
are focused on major topics. Opinion leaders and other users classified as
``evangelists'' contribute to diffuse major as well as minor topics to audiences
far from the core of the network, and normal users usually have a low
contribution on the diffusion process, and are more passive
consumers than active producers of contents.

In social networks, and in Twitter in particular, different sources of
information diffusion coexist. In fact, information may come and is propagated directly from
  the users within the platform following the word-of-mouth effect~\cite{Goldenberg2001Talk},
  as well as from other sources that are external to social networks (e.g. television and radio)~\cite{Myers2012}.
Moreover, analyses of information cascades in Twitter confirm that
  various elements impact on the information diffusion
  process~\cite{Arnaboldi2014Information,pezzoni2013}.  Among others, the
  ``standing'' of users (e.g., their importance in their personal social
  network), their network centrality (according to standard complex network
indices), as well as the freshness of information are directly related to the
probability of propagation across nodes, and thus ultimately impact on the
breadth of information cascades. In this work, we are interested in
characterising in detail information diffusion through the word-of-mouth effect.
In particular, we focus on the analysis of information diffusion
seen from an ego network perspective, and we assess the impact of ego network
circles on the process. To the best of our knowledge, this is the first detailed
analysis on these particular aspects.

Before describing the analysis to characterise the structural properties of
OSNs, we present a description of the Facebook and Twitter data sets
that we used.

\section{Online Communications Data Sets}
\label{sec:datasets}

To study the
structural properties of OSNs and to assess their role in the diffusion of
information in the network, we have analysed two data sets containing traces of
communication between people in Facebook and Twitter, two amongst the most
important social media nowadays (see ~\ref{app:platforms} for a
detailed description of these platforms).  From
the data sets, we have obtained the frequency of contact between online users,
that has been used to estimate the strength of the social links (as we
discuss in Section~\ref{sec:freqcont}). Hence, we have built an ego network for each
    user, and we have analysed their structural properties (in
    Section~\ref{sec:analysis}) and their role in the diffusion of information
    (in Section~\ref{sec:diffusion_analysis}).

\subsection{Data Download} \label{sec:download}

\subsubsection{Facebook}
Although Facebook generates a huge amount of data regarding social
communications between people, obtaining these data is not easy. In fact,
publicly available data have been strongly limited by the introduction of strict
privacy policies and default settings for the users after 2009. Nevertheless,
before that date, most of the user profiles were public and the presence of the
\emph{network} feature, that has been removed in 2009, allowed researchers to
collect large-scale data sets containing social activity between users. A
network was a membership-based group of users with some properties in common
(e.g. workmates, classmates or people living in the same geographical region).
Each user profile was associated to a regional network based on her geographical
location. By default, each user of a regional network allowed other users in the
same network to access her personal information, as well as her status updates
and the posts and comments that she received from her friends. Exploiting
these
characteristics of regional networks, some data sets have been downloaded, such
as those described in~\cite{Wilson2012}, which have been made partly publicly available for
research\footnote{http://current.cs.ucsb.edu/facebook/}. In this paper, we used the
data set referred as ``Regional Network A''.

The use of the regional networks feature allowed researchers to download large
data sets from Facebook, however, it entails some limitations that must be taken
into account for our analysis. In fact, the considered data set contains
information regarding only the users and their social
interactions within a regional network, excluding all the interactions and the social links that
involve users external to this area. Therefore, assuming that, for each user, a
part of her social relationships involve people who do not belong to the same
network, this could lead to a reduction of the ego networks' size. Moreover, we
do not have specific information about the completeness of the crawling process
that should have downloaded only a sample of the original regional network. For
example, in~\cite{Wilson2012} the same crawling agent was used for downloading
several other regional networks (not publicly available) collecting, on average,
$56.3\%$ of the nodes and $43.3\%$ of the links. We used this additional
  knowledge in the analysis to obtain the highest possible accuracy in the results,
as explained in detail in Section~\ref{sec:analysis}.

\subsubsection{Twitter}

For Twitter, we have implemented a
crawling agent that is able to download user profiles and their communication
data from Twitter. The agent visited the Twitter graph considering the users as
nodes and following the links between them. In our study, a link between two
nodes exists if at least one of the users follows the other or an interaction
between them has occurred. As an indication of an interaction, we use the presence
of a \emph{mention} in a tweet (i.e. the fact that a user explicitly mentions
the other in a tweet) and a \emph{reply} (i.e. a direct response to a tweet).

The crawling agent starts from a given user profile (seed) and visits the
Twitter graph following the links. For each visited node, we took advantage of
the Twitter REST API to extract the user \emph{timeline} (i.e. the list of
posted tweets that can include mentions and replies), the \emph{friends} list
(i.e. the people followed by the user) and the \emph{followers} list (i.e. the
people who follow the user). Twitter REST API limits the amount of tweets that
can be downloaded per user up to $3,200$ tweets. This does not represent a
constraint to our analysis since, as we show in the following, it is sufficient
for our purposes.

\begin{table}[t]
  \renewcommand{\arraystretch}{1.2}
  \caption{Statistics of the Facebook social graph} \label{tab:sgstat}
  \centering
  \begin{tabular}{|l|c|}
    \hline
    \# Nodes & $3,097,165$\\
    \# Edges & $23,667,394$\\
    Average degree & $15.283$\\
    Average shortest path & $6.181$\\
    Clustering coefficient & $0.209$\\
    Assortativity & $0.048$\\
    \hline
  \end{tabular}
\end{table}

The crawling agent uses $250$ threads that concurrently access a single queue
containing the IDs of the user profiles to download. Each thread extracts a
certain number of user IDs from the queue,
then it gets the related profiles and
communication data from Twitter using the REST API. Finally, after extracting
new user IDs from the communication data and from
the friends/follower lists,
the threads add them to the queue. The use of multiple threads allowed
us both to
speed-up the data collection and to avoid that the crawler
remains trapped in
visiting the neighbourhood of a node with a large number of links. The seed that we
used to start the data collection is the profile of a widely know user (user
id:~$813286$), so that her followers represent an almost random sample of the
network.

The crawling agent allowed us to obtain a snowball sample of a complete portion
of the Twitter network. Compared to the Facebook data set, this contains complete
ego networks.

\subsection{Data Sets Properties}

\subsubsection{Facebook}

The Facebook data set that we used in this work consists of a \emph{social
graph} and four \emph{interaction graphs}. These graphs are defined by lists of
edges connecting pairs of anonymised Facebook user IDs.

The social graph describes the overall structure of the downloaded network. It
consists of more than 3 million nodes (Facebook users) and more than $23$
million edges (social links). An edge represents the mere existence of a
Facebook friendship, regardless of the quality and the quantity of the
interactions between the involved users. Basic statistics\footnote{The
  clustering coefficient is calculated as the average local clustering
  coefficient (Equation 6 in~\cite{newman2003structure}).} of the social graph
  are reported in Table~\ref{tab:sgstat}.

\begin{table}[t]
  \renewcommand{\arraystretch}{1.2}
  \scriptsize
  \caption{Statistics of the Facebook interaction graphs (preprocessed).}
  \label{tab:igstat}
  \centering
  \begin{tabular}{l|cccc|}
    \cline{2-5} &
    \textbf{Last mo.} & \textbf{Last 6 mo.} & \textbf{Last year} &
    \textbf{All}\\
    \hline
    \multicolumn{1}{|c|}{\# Nodes} & $414,872$ & $916,162$
    & $1,133,151$ & $1,171,208$\\ \multicolumn{1}{|c|}{\# Edges} & $671,613$ &
    $2,572,520$ & $4,275,219$ & $4,357,660$\\
    \multicolumn{1}{|c|}{Avg. degree}
    & $3.238$ & $5.616$ & $7.546$ & $7.441$\\
    \multicolumn{1}{|c|}{Avg. weight}
    & $1.897$ & $2.711$ & $3.700$ & $3.794$\\
    \hline
  \end{tabular}
\end{table}
  
The social graph can be used to study the global properties of the network, but
alone it is not enough to make a detailed analysis of the structure of social
ego networks in Facebook. Indeed, this analysis requires an estimation of the
strength of the social relationships. To this aim, in
Section~\ref{sec:freqcont}, we leverage the data contained in the interaction
graphs to extract the frequency of contact of the social links that can be used
to estimate the tie strength.

Interaction graphs describe the structure of the network during specific
temporal windows, providing also the number of interactions occurred for each
social link. The four temporal windows in the data set, with reference to the
time of the download, are: \emph{last month}, \emph{last six months}, \emph{last
year} and \emph{all}. The latter temporal window (``all'') refers to the whole
period elapsed since the establishment of each social link, thus considering all
the interactions occurred between the users. In an interaction graph, an edge
connects two nodes only if an interaction between two users occurred at least
once in the considered temporal window. The data set that we used for the
analysis contains interactions that are either Facebook Wall posts or photo
comments.

In Facebook, an interaction can occur exclusively between two users who are
friends. In other words, if a link between two nodes exists in an interaction
graph, an edge between the same nodes should be present in the social graph.
Actually, the data set contains a few interactions between users which do
not correspond to any link in the social graph. These interactions probably refer to expired
relationships or to interactions made by accounts that are no longer active. To
maintain consistency in the data set, we excluded these interactions from
the analysis. The amount of discarded links is, on average, $6.5\%$ of the total
number of links in the data set.

In Table~\ref{tab:igstat}, we report some statistics regarding the different
interaction graphs. Each column of the table refers to an interaction graph
related to a specific temporal window. The average degree of the nodes is the
average number of social links per ego that have at least one interaction in
the considered temporal window. Similarly, the average edge weight represents
the average number of interactions for each social link.

\subsubsection{Twitter} \label{sec:dsproptwitter}

We collected a data set from $2,463,692$ Twitter users, whose data were
downloaded between November 2012 and March 2013. In contrast to Facebook, whose
users are generally people who want to socialise with others, communicating and
maintaining social relationships, Twitter users are more heterogeneous. In fact,
the downloaded accounts can also be related to companies, public figures, news
broadcasters, bloggers and many others. We can thus classify the users in two
different categories: (i) \emph{socially relevant users}, which represent people
who use Twitter for socialising, and (ii) \emph{other users}, which use
Twitter for other purposes. This classification is fundamental for our
study since, in order to analyse the human social behaviour, we want to consider
socially relevant users only. To automatically distinguish between the
two classes of users, we built a classifier based on
Support Vector Machines (SVM) that, relying on the activity logs and on the
meta-data of the accounts in the data sets, distinguishes socially relevant
users from other users. The accuracy of the SVM is 83\%, and the false
positives rate around 8\%. The details of the classifier are described in
\ref{app:classifier}. Note that, also in Facebook, some accounts
represent users that are not socially relevant (e.g. companies and public
figures). Nevertheless, Facebook is more naturally used as a private
communication channel, and public communications (e.g. status updates) are not
considered in the data set. For this reason, and for the lack of sufficiently
detailed information about the nature of Facebook users in the data set, we
analysed all the Facebook accounts without splitting them into separate classes.

\begin{table}[t]
  \renewcommand{\arraystretch}{1.2}
  \caption{Twitter data set
  (all users) and classes statistics.} \label{tab:data_set}
  \scriptsize
  \begin{tabular}{c|ccc|}
    \cline{2-4} & \textbf{All users} & \textbf{Soc. rel.
    users} & \textbf{Other users}\\
    \hline
    \multicolumn{1}{|c|}{$N$} &
    $2.463.692$ & $1,653,436$ & $810,256$ \\
    \multicolumn{1}{|c|}{$N_{3,200}$} &
    $510,119$ & $260,632$ & $249,487$ \\
    \multicolumn{1}{|c|}{$(\%~N_{3,200})$}
    & $(20.7\%)$ & $(15.8\%)$ & $(30.8\%)$ \\
    \hline
    \multicolumn{1}{|c|}{$\#~Tweets$} & $1,207$ & $979$ & $1,696$ \\
    \multicolumn{1}{|c|}{$\#~Following$} & $3,157$ & $2,553$ & $4,448$ \\
    \multicolumn{1}{|c|}{$\#~Followers$} & $7,353$ & $2,744$ & $17,201$ \\
    \hline \multicolumn{1}{|c|}{$\%~Tweets_{\mathrm{REPL}}$} & $17.4\%$ &
    $18.4\%$ & $15.4\%$ \\
    \multicolumn{1}{|c|}{$\%~Tweets_{\mathrm{MENT}}$} &
    $22.7\%$ & $21.6\%$ & $24.7\%$ \\
    \hline
  \end{tabular}
\end{table}

In the column ``all users'' of Table~\ref{tab:data_set}, we present some
statistics of all the users in the data set, while in the next two columns we
present the statistics of the socially relevant users and of the other users
respectively. For each category, we present the number of users $N$ and the
average number of tweets, friends, and followers. Each average value is reported
with its $95\%$ confidence interval between square brackets.

We can notice that socially relevant users are the majority and their statistics
indicate that they are less active than the other users. This could be explained
by the fact that users in the ``other users'' class may be companies or other
kinds of accounts managed by more than one person at the same time and aimed at
advertising goods or services.

In the table, we also report, for each class of users, the average ratio of
replies ($tweets_{\mathrm{REPL}}$) and mentions ($tweets_{\mathrm{MENT}}$),
calculated over the total number of tweets. These values indicate that around
$40\%$ of the tweets downloaded by our crawler contain mentions or replies
between people. These tweets are important for our study since they represent
direct interactions, rather than broadcast communications. Moreover, socially
relevant users show a slightly higher percentage of replies than other types of
users ($18.4\%$ vs. $15.4\%$), indicating that they use more directional
communications, a typical human social behaviour.

\begin{figure}[t]
  \centering
  \includegraphics[width=0.4\textwidth]{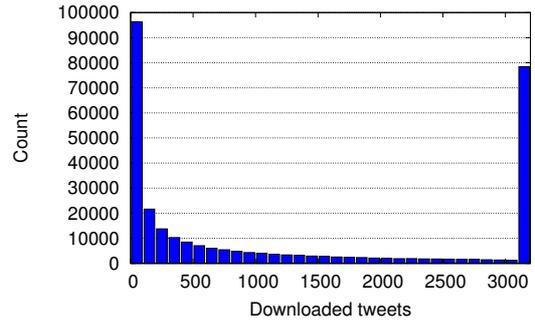}
  \caption{Downloaded tweets per user distribution.}
  \label{fig:tweet_count_hist}
\end{figure}

In Figure~\ref{fig:tweet_count_hist}, we show the distribution of the number of
tweets downloaded per user. We can notice the presence of a peak corresponding
to the value $3,200$, which is the maximum amount of tweets downloadable using the
Twitter REST API. Cases where the number of tweets is lower than $3,200$
correspond to users that have generated less than $3,200$ tweets from the
  creation of their accounts. The number of users that posted a number of tweets
above this threshold is indicated in the table by $N_{3,200}$. Note that, for
socially relevant users, this is a relatively small fraction of the total number
of users ($15.8\%$). This means that our crawler was able to download the
entire twitting activity for the majority of the users relevant for our study,
and for those users for whom we have not obtained the entire history of outgoing
communications, we still have a significant number of tweets.

In order to further investigate the behavioural differences between socially
relevant users and the other users, we studied the number of replies the
users send to their friends on average. In~\cite{Goncalves2011pone}, a similar
analysis was used to conclude that a concept similar to the Dunbar's number
(the maximum number of active social relationships an individual can actively
maintain) holds also in Twitter.

\begin{figure}[t]
  \centering
  \includegraphics[width=0.4\textwidth]{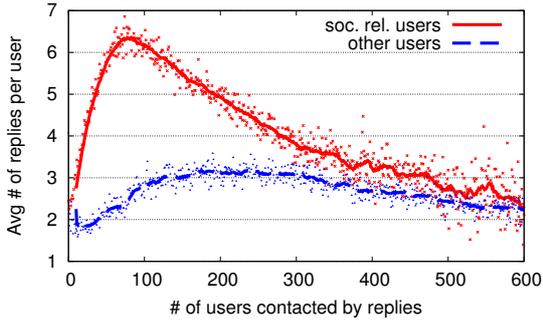}
  \caption{Points
  represent the average number of replies made by accounts with different number
of friends; thick lines are their running averages.}
\label{fig:vespignani}
\end{figure}

Figure~\ref{fig:vespignani} depicts the trend of the average number of replies
per friend as a function of the number of friends of the user. Differently
from~\cite{Goncalves2011pone}, we have divided the analysis for the two classes
identified: ``socially relevant users'' and ``other users''. The results highlight a clear distinction between the properties of
the two classes.

Socially relevant users show a higher mean value of replies per friend and a
maximum around $80$ friends. This is an indication of the effect of the
cognitive limits of human brain on the ability to maintain social relationships
in OSNs. The peak of the curve identifies the threshold beyond which the effort
dedicated to each social relationship decreases. This is due to the exhaustion
of the available cognitive/time resources, which, therefore, have to be split over
an increasing number of friends. As discussed in~\cite{Goncalves2011pone}, this
can be seen as an evidence of the presence of the Dunbar's number in
Twitter.

Other users show a quite different pattern,
with lower average value of replies per
friend without any significant discontinuities. This indicates that accounts
belonging to the class ``other users'' are not influenced by cognitive
capabilities. In fact they, are often managed by more than one person or by
non-human agents.

\subsection{Obtaining the Frequencies of Contact} \label{sec:freqcont}

\subsubsection{Facebook}

In order to characterise tie strength in Facebook, we need to estimate the
\textit{link duration}, that is the time elapsed since the establishment of the
social link. This is essential to calculate the frequency of contact
between the users involved in a social link, and the latter is then
used to estimate the tie
strength. In the literature, the duration of a social link is commonly estimated
using the time elapsed since the first interaction between the involved
users~\cite{Gilbert2009}. Unfortunately, the data set does not provide any
indication regarding the time at which the interactions occurred. To overcome
this limitation, we have approximated the links duration leveraging the
difference between the number of interactions made in the different temporal
windows. Details on how we have estimated the link duration and the frequency of
contact between users in the Facebook data set are given in
\ref{app:facebook}. The frequency of contact between pairs of users has
been calculated as the total number of interactions occurred (obtained from the
``all'' interaction graph) divided by the estimated duration of their social
link. In case the users have never interacted their frequency of contact is set
to zero.

\subsubsection{Twitter}

The Twitter data set contains all the tweets sent by the users (with the limit
of $3,200$ tweets per user). Hence, obtaining the frequency of contact between
users in Twitter is more straightforward than in Facebook. Considering socially
relevant users with all their social contacts, we calculated the duration of each social
link as the time elapsed between the first mention or reply exchanged between
the involved users and the time of the download. Given a social link, we have
thus calculated the frequency of contact for each of the two users as the number
of replies sent to the other divided by the duration of the social link. In the
calculation, we have used the number of replies since it is the strongest
indicator of the strength of a social link in Twitter and since it has been
already used in previous work~\cite{Goncalves2011pone}.

\section{Ego Networks Structure in Online Social Networks} \label{sec:analysis}

In this section, we analyse the structures of the ego networks that can be
identified in Facebook and Twitter and we compare them with the model for
offline social networks presented in Section~\ref{sec:microlevel}. 

In order to extract the ego networks from our data sets, we have grouped the
relationships of each user into different sets\footnote{Since social links in
the Facebook interaction graphs represent undirected edges, we have duplicated
each social link in the data set in order to consider it in both the ego
networks of the users connected by it.}. Then, to avoid including possible
outliers in the analysis, we have selected only the ego networks that meet the
following criteria:

\begin{enumerate} \item \emph{The account of the ego must have been created at
      least six months before the time of the download}. In case of the Facebook
      data set, the lifetime of the accounts is estimated as the time since the
      user made the first interaction. In case of the Twitter data set, we know
      the time of the account creation as it is included in the meta-data we
      downloaded.  \item \emph{Ego must have made, on average, $10$ or more
	interactions per month}. For both data sets, we can calculate the
	average activity as the total number of registered interactions divided
	by the lifetime of the account.  \end{enumerate}

This selection is also motivated by the findings in other OSNs analyses (see for
example~\cite{zhao2012multi}), in which ego networks are found to be highly
unstable and with a high growing rate soon after ego joins the network, but tend
to be stable after the first few months of activity. This selection allowed us
to consider only users who regularly use OSNs, and filter out typical initial
bursts of activities of new users. This resulted in the selection of $91,347$
ego networks from the Facebook data set and $394,238$ ego networks from the
Twitter data set. These numbers, as we will see later, are sufficient to draw
significant results about the ego network properties of OSNs. Note that the
selected socially relevant users can have both socially relevant users and other
users in their ego networks. In our analysis, we consider all the possible kinds
of alters of socially relevant users. This is important to have a complete view
of the structure of their social networks, since each ego spends cognitive
efforts for communicating with all her alters, and the properties of her ego
network are impacted by her cognitive and time constraints, no matter whether
she spends all her time communicating with robots or with other humans.

\subsection{Analysis of the Aggregated Frequency Distribution}

\begin{figure}[t]
  \subfloat[Facebook]{%
    \includegraphics[width=0.4\textwidth]{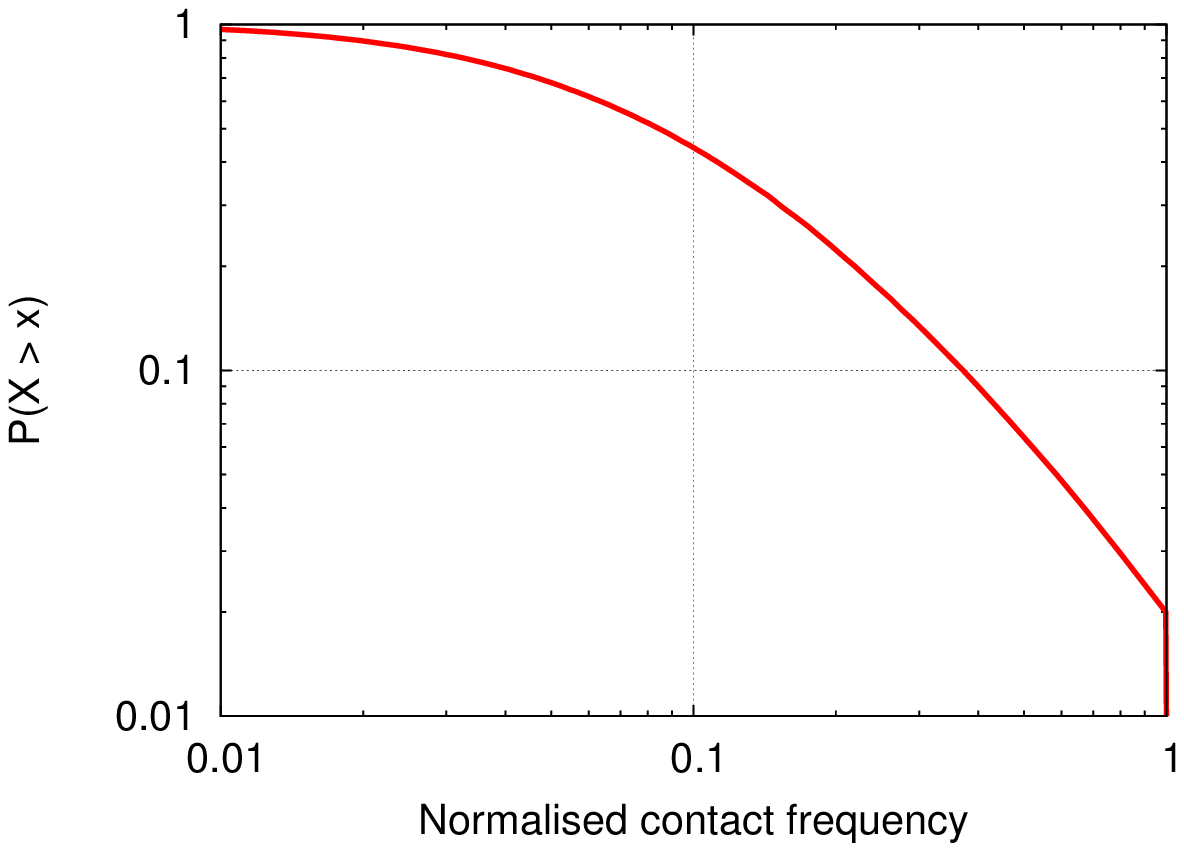} }
    \hfill
    \subfloat[Twitter\label{subfig-2:twitter_normfreq}]{%
      \includegraphics[width=0.4\textwidth]{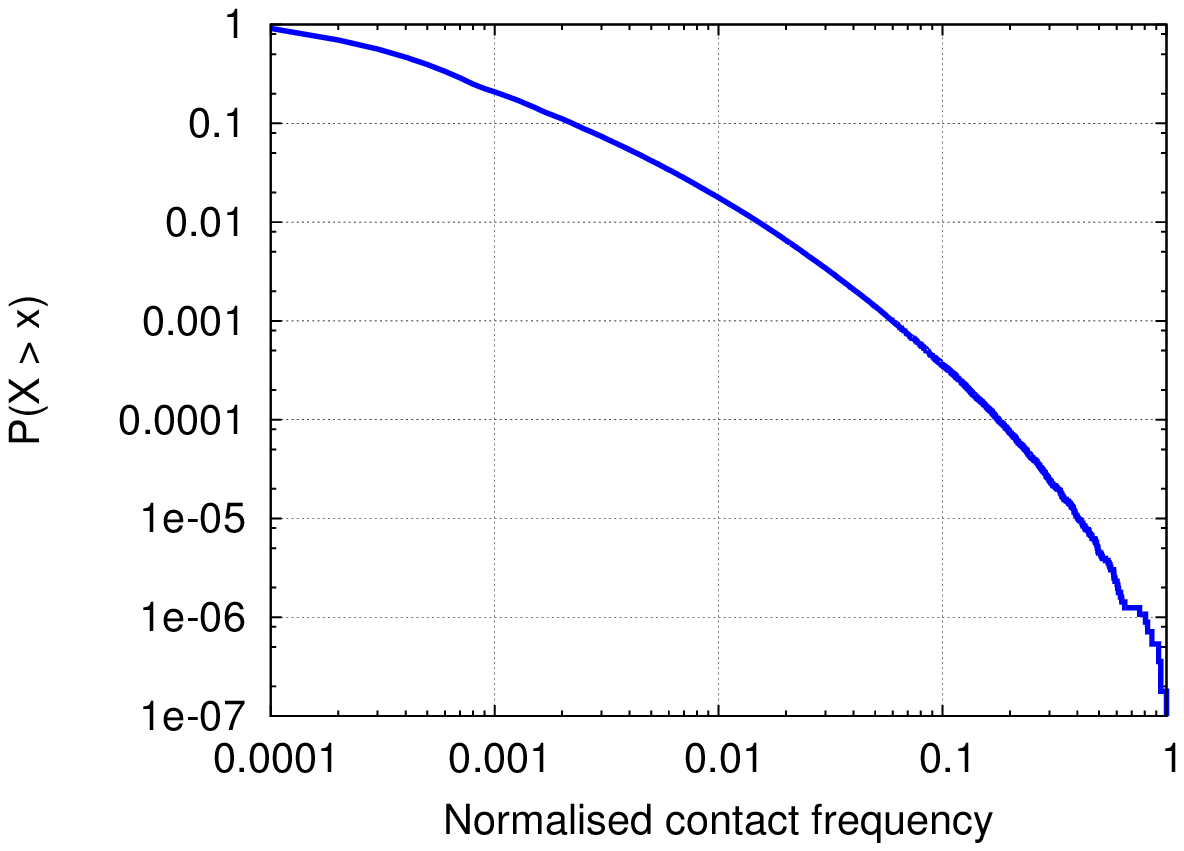}
    }
    \caption{Aggregated CCDF of the normalised frequency of contact for all
      the ego networks in the data sets.}
      \label{fig:ccdf_freqnorm}
\end{figure}

The possible presence of social structures in Facebook and Twitter may be
revealed by steps in the distribution of the frequency of contact since it is
the key aspect to quantify the tie strength. If the frequency of contact of an
ego network gracefully degrades and does not present steps in the distribution,
this suggests the absence of any structure. On the contrary, if the frequency of
contact appears clustered in different intervals, each of them may reveal the
presence of a ego network layer.

A simple initial analysis to check the presence of such steps in the
distribution is considering the CCDF of the aggregate normalised frequency of
interaction. In particular, we have considered the distribution obtained by
taking together all the frequencies of contact of all ego networks in each data
set. A normalisation of the frequencies of contact for each ego network is
necessary in order to level out the differences between users in the use of the
platforms. Analysing the aggregate distribution permits to focus on a single
distribution, instead of analysing all individual ego networks' distributions.
The obtained CCDFs, depicted in Figure~\ref{fig:ccdf_freqnorm}, show a smooth
trend. Clearly, this does not allow us to conclude that ego networks are
clustered, but is not a sufficient condition to rule out this hypothesis. In
fact, even if the individual ego network distribution had a social structure,
and therefore steps in their distributions would be present, such steps
may appear at different
positions from one network to another, thus resulting in a smooth aggregate CCDF
(remember that also in the ego network model the sizes of the layers are average
values, but variations are possible at an individual ego network
level).

The CCDFs show a long tail, which can be ascribed to a power law shape. A power law shape in the
aggregate CCDF is a necessary condition to have power law distributions
in at least one
ego network~\cite{Boldrini2011}. However, this is not a sufficient condition to
have power law distributions in each single CCDF~\cite{Conti2011a}.
Although formally the presence
of a long tail in the CCDF is not a conclusive proof of the existence of small
numbers of very active social links in the individual ego networks, this
is anyway a strong indication in this direction, and a possible similarity
between ego networks in offline and online social
networks. Studies in the social and anthropology literature revealed that ego
networks are
characterised by a small set of links with very high frequencies of contact
(corresponding to the links in the support clique), which appear as a heavy tail
in the CCDF of individual ego network contact frequency.

\subsection{Revealing Ego Network Structure through Clustering}
\label{sec:clustanalysis}

%\begin{figure}[t] \centering
%\includegraphics[width=0.4\textwidth]{singleCCDFex.eps} \caption{CCDF of the
%normalised frequency of contact of an individual Twitter ego network.}
%\label{fig:ccdfex} \end{figure}

%As shown in Figure~\ref{fig:ccdfex}, the CCDF distributions of individual ego
%networks present a series of steps that were hidden in the aggregate
%distribution analysed in the previous section. As previously said, the presence
%of these steps reveals the underlying ego network structure.

To further investigate the online ego network structures, we have applied
cluster analysis on the normalised frequencies of contact of each ego network,
looking for the emergence of layered structures. For each ego network, the
frequencies of contact between ego and alters represent a set of values in a
mono-dimensional space. Applying cluster analysis to mono-dimensional values
does not require advanced clustering techniques, therefore we can consider
standard widely-used methods such as \emph{$k$-means clustering} and
\emph{density-based clustering} (e.g. DBSCAN algorithm). Using $k$-means
clustering, given a fixed number of clusters $k$, the data space is partitioned
so that the sum of squared euclidean distance between the centre of each cluster
(centroid) and the objects inside that cluster is minimised. In density-based
clustering, clusters are defined as areas of higher density than the remainder
of the data set, which is usually considered to be noise~\cite{Kriegel2011}.
In~\cite{Arnaboldi2012}, both clustering techniques have been applied on the same
Facebook data set used in the present analysis. Nonetheless, results showed that
the clusters identified by the two methods are substantially
equivalent and that both can be used for the study of social structures in ego
networks leading to the same conclusions~\cite{Arnaboldi2012}.

In this work, we report the analysis using the $k$-means clustering since it is
the simplest and the most computationally affordable method. This method is
defined as an optimisation problem that is known to be NP-hard. Because of this,
the common approach for $k$-means clustering is to search only for approximate
solutions. Fortunately, in the special case of mono-dimensional space, we can
use an algorithm, called \texttt{Ckmeans.1d.dp}, able to always find the optimal
solution efficiently~\cite{Wang2011}.

%In Figure~\ref{fig:ccdfex} we show the result of the \texttt{Ckmeans.1d.dp}
%algorithm (with $k = 4$) applied to the frequencies of contact of an individual
%Twitter ego network. As expected, the limits between adjacent clusters (red
%bars in the figure) are placed by the algorithm in correspondence of the steps
%in the CCDF distribution.

\subsubsection{Typical Number of Clusters}

\begin{figure}[t]
  \centering
  \includegraphics[width=0.4\textwidth]{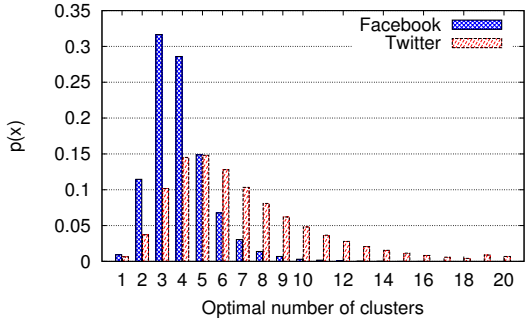}
  \caption{Density function of $k^*$ in Facebook and Twitter ego networks.}
  \label{fig:kopt}
\end{figure}

In the first step of our cluster analysis, we have sought, for each ego network,
the typical number of clusters (i.e. the number $k^*$) in which the frequencies
of contact can be naturally partitioned. In order to do this, we have evaluated
the goodness of the result of different clustering configurations. For $k$-means
methods, this is usually expressed in terms of \emph{explained variance}.
In fact, a small variance in the individual clusters
means that data are well
described by the current clustering,
and this is evidenced by a high value of
the explained variance (up to the maximum value $1.0$). Specifically, the
explained variance is defined by the following formula:

\begin{equation}
  VAR_{exp}=\frac{SS_{tot}-\sum_{j=1}^{k}{SS_j}}{SS_{tot}},
  \label{eq:expvar}
\end{equation}
where $j$ is the $j^{th}$ cluster, $SS_j$ is the sum of squared distances within
cluster $j$ and $SS_{tot}$ is the sum of squared distances of the all the values
in the data space. Given a vector $\mathrm{\mathbf{X}}$, the sum of squared
distances $SS_{\mathrm{\mathbf{X}}}$ is defined as
$SS_{\mathrm{\mathbf{X}}}=\sum_{i}{(x_i-\mu_{\mathrm{\mathbf{X}}})^2}$, where
$\mu_{\mathrm{\mathbf{X}}}$ denotes the mean value of $\mathrm{\mathbf{X}}$.

Given the number of clusters $k$, $k$-means clustering algorithms partition the
space minimising the sum of squared distance within the clusters
$\sum_{j=1}^{k}{SS_j}$. According to Equation~\ref{eq:expvar}, for a given $k$,
  the solution of $k$-means clustering also provides the maximum value of the explained
variance $VAR_{exp}$, since the sum of squared distances $SS_{tot}$ is constant
given the data space. In principle, the optimal number of clusters $k^*$
  would be  equal to the number of objects in the data space, as the value of
  $VAR_{exp}$ increases monotonically with $k$. Thus, there is a inherent overfitting problem. To overcome this
problem and determine the typical number of clusters we used the Akaike
Information Criterion (AIC), an information-theoretic measure that trades off
distortion against model complexity, defined by the following
equation:

\begin{equation}
  AIC=-2L(k) + 2q(k)
  \label{eq:AIC}
\end{equation}
where $-L(k)$ is the negative maximum log-likelihood of the data for $k$ clusters,
and is a measure of distortion. $q(k)$ is the number of parameters of the model
with $k$ clusters and measures complexity. The model showing the minimum value
of $AIC$ is the one with the best trade-off between distortion and complexity.

We have calculated the AIC for all the ego networks in Facebook and Twitter, by
applying $k$-means with $k$ from $1$ to $20$. For each ego network we define as
$k^*$ the the value of $k$ that minimises equation~\ref{eq:AIC}. In
Figure~\ref{fig:kopt}, we
report the density function of $k^*$ for the ego networks in our data sets.

\begin{table*}[t] \caption{Optimal number of clusters ($k^*$) of ego networks.}
  \label{tab:kopt} \renewcommand{\arraystretch}{1.2} \footnotesize \centering
  \begin{tabular}{c|cc|cc|} \cline{2-5} &
    \multicolumn{2}{|c|}{\textbf{Facebook}} &
    \multicolumn{2}{|c|}{\textbf{Twitter}} \\ \hline
    \multicolumn{1}{|c|}{\textbf{$\mathrm{k_{opt}}$}} & \textbf{\# of nets} &
    \textbf{Net size} & \textbf{\# of nets} & \textbf{Net
    size}\\
    \hline
    \multicolumn{1}{|c|}{$1$} & $844$
    $(0.9\%)$ & $29.68$ $[\pm1.95]$ &             $2,500$ $(0.6\%)$ &
    $192.77$ $[\pm12.44]$\\
    \multicolumn{1}{|c|}{$2$} & $10,465$
    $(11.47\%)$ & $41.82$ $[\pm0.39]$ &            $14,683$ $(3.7\%)$
    & $104.93$ $[\pm2.35]$\\
    \multicolumn{1}{|c|}{$3$} & $28,918$
    $(31.66\%)$ & $39.00$ $[\pm0.27]$ &       $40,099$ $(10.2\%)$ &
    $91.42$ $[\pm0.98]$\\
    \multicolumn{1}{|c|}{$4$} & $26,124$
    $(28.60\%)$ & $41.99$ $[\pm0.38]$ &      $57,227$ $(14.5\%)$ &
    $89.09$ $[\pm0.73]$\\
    \multicolumn{1}{|c|}{$5$} & $13,584$
    $(14.87\%)$ & $53.89$ $[\pm0.66]$ &         $58,410$ $(14.82\%)$ &
    $92.56$ $[\pm0.70]$\\
    \multicolumn{1}{|c|}{$>5$} & $11,412$
    $(12.50\%)$ & $82.02$ $[\pm1.00]$ &       $221,319$ $(56.1\%)$ &
    $100.42$ $[\pm0.31]$\\
    \hline
  \end{tabular}
\end{table*}

We have found that the distribution of $k^*$ has a peak between 3 and
4 for Facebook
and between $4$ and $5$ for Twitter. The presence of a typical number of
clusters close to $4$ is the first indication of similarity between the findings
in offline and online ego networks.

In Table~\ref{tab:kopt}, we report the properties of the ego networks found with
different numbers of $k^*$. The average network size (``net size'' in the table)
is reported with its $95\%$ confidence interval between square brackets.

Ego networks with only one circle tend to have similar values of contact
frequency for all their links, and in many cases the contact
frequencies are
exactly the same. This could be ascribed to automated forwarding of messages on
all the links, associated to bots or spammers, and indicates the presence of a
small set of biased ego networks in the data set. Remember that, although the
classifier we used to select socially relevant users has a high accuracy, some
accounts could be false positives, as probably in this case. Whilst the size
of the ego networks with one circle in Facebook is relatively small, in Twitter
we notice very large ego networks (i.e. with average size of $192.77$ alters).
This could be explained by the fact that it is more difficult for bots
or spammers
to create a large network of social relationships in Facebook, whereas in
Twitter is easier to have a large number of followers with a significant
interaction. This is due to the
differences in the nature of the two platforms. In fact, in
Facebook users tend
to accept friendships requests only if they know the requester in person, or
they recognise a real human behind her profile, whilst in Twitter the
heterogeneity of profiles makes this kind of selection more difficult.

The size of the ego networks seems to be almost
constant between two and five circles, and it increases for networks with more
than five circles.

\subsubsection{Ego Network Circles}

\begin{table*}[t] \caption{Ego network circles' properties.}
  \label{tab:clusters} \setlength{\tabcolsep}{10pt}
  \renewcommand{\arraystretch}{1.2} \footnotesize \centering
  \begin{tabular}{cl|cccccccccc|} \cline{3-12} & &
    \multicolumn{2}{c}{\textbf{$\mathrm{C_1}$}} &
    \multicolumn{2}{c}{\textbf{$\mathrm{C_2}$}} &
    \multicolumn{2}{c}{\textbf{$\mathrm{C_3}$}} &
    \multicolumn{2}{c}{\textbf{$\mathrm{C_4}$}} &
    \multicolumn{2}{c|}{\textbf{$\mathrm{C_5}$}}\\
    \hline
    \multicolumn{1}{|c|}{\multirow{3}{*}{\textbf{Facebook}}} & min freq. &
    \multicolumn{2}{c}{$5.09$} & \multicolumn{2}{c}{$1.95$} &
    \multicolumn{2}{c}{$0.67$} & \multicolumn{2}{c}{$0.11$} &
    \multicolumn{2}{c|}{$-$}\\ \multicolumn{1}{|c|}{} &
    size\tablefootnote{Facebook circles' size are affected by the data set
      subsampling discussed in Section~\ref{sec:download}.} &
      \multicolumn{2}{c}{$(1.79)$} & \multicolumn{2}{c}{$(5.83)$} &
      \multicolumn{2}{c}{$(17.05)$} & \multicolumn{2}{c}{$(50.46)$} &
      \multicolumn{2}{c|}{$-$}\\ \multicolumn{1}{|c|}{} & scal. fact. & &
      \multicolumn{2}{c}{~~$3.26$} & \multicolumn{2}{c}{~~~$2.93$} &
      \multicolumn{2}{c}{~~~~$2.96$} & \multicolumn{2}{c}{~~~~$-$} & \\
      \hline
\multicolumn{1}{|c|}{\multirow{3}{*}{\textbf{Twitter}}} & min freq. &
\multicolumn{2}{c}{$20.55$} & \multicolumn{2}{c}{$8.91$} &
\multicolumn{2}{c}{$3.98$} & \multicolumn{2}{c}{$1.36$} &
\multicolumn{2}{c|}{$0.18$}\\
\multicolumn{1}{|c|}{} & size &
\multicolumn{2}{c}{$1.66$} & \multicolumn{2}{c}{$5.06$} &
\multicolumn{2}{c}{$12.87$} & \multicolumn{2}{c}{$32.66$} &
\multicolumn{2}{c|}{$97.47$}\\
\multicolumn{1}{|c|}{} & scal. fact. & &
\multicolumn{2}{c}{~~$3.04$} & \multicolumn{2}{c}{~~~$2.55$} &
\multicolumn{2}{c}{~~~~$2.54$} & \multicolumn{2}{c}{~~~~$2.98$} &\\
\hline
\multicolumn{1}{|c|}{\multirow{3}{*}{\textbf{Offline}}} & min freq. &
\multicolumn{2}{c}{$4.29$} & \multicolumn{2}{c}{$1.00$} &
\multicolumn{2}{c}{$-$} & \multicolumn{2}{c}{$0.08$} & \multicolumn{2}{c|}{$-$}\\
\multicolumn{1}{|c|}{} & size & \multicolumn{2}{c}{$4.6$} &
\multicolumn{2}{c}{$14.3$} & \multicolumn{2}{c}{$42.6$} &
\multicolumn{2}{c}{$132.5$} & \multicolumn{2}{c|}{$-$}\\
\multicolumn{1}{|c|}{}
& scal. fact. & & \multicolumn{2}{c}{~~$3.10$} & \multicolumn{2}{c}{~~~$2.98$} &
\multicolumn{2}{c}{~~~~$3.11$} & \multicolumn{2}{c}{~~~~$-$} &\\
\hline
\end{tabular}
\end{table*}

According to the previous analysis, the typical number of clusters in online ego
networks appears to be equal to $3-4$ in Facebook and $4-5$ in Twitter. Yet, to
be able to compare the structure of online ego networks with that
found in
offline networks we have applied the algorithm \texttt{Ckmeans.1d.dp} with $k=4$
for Facebook and $k=5$ for Twitter. This choice will be more clear in the
following, but we motivate it anticipating that in Twitter a new internal
circles appear, that is not visible in the used Facebook
  data set. For each ego network, we obtained a set of clusters
that we refer as $S_1$, $S_2$, $S_3$, $S_4$, and
$S_5$ (where needed), sorted by decreasing value of the centroid (i.e. the
average frequency of contact of the cluster) so that $S_1$ represents the
cluster of the social links with the highest frequency of contact. The obtained
clusters are not directly comparable with the circles of offline ego networks
discussed in Section~\ref{sec:background}. In fact, while clusters are disjoint
groups, social circles, as depicted in Figure~\ref{fig:egonet}, are
hierarchically inclusive (i.e. the \emph{support clique} is included in the
\emph{sympathy group} which is included in the \emph{affinity group} which is
included in the \emph{active network}). For this reason, in order to compare
social structures in online and offline ego networks, we have aggregated the
clusters to form hierarchically inclusive circles. Specifically, we have defined
the circles $C_1$, $C_2$, $C_3$, $C_4$, and $C_5$ as $C_k = \bigcup_{i=1}^k S_i$
so that $C_1 \subseteq C_2 \subseteq C_3 \subseteq C_4 \subseteq C_5$.

In Table~\ref{tab:clusters}, we compare the properties of the circles in Facebook
and Twitter ego networks with those found in offline ego networks. One of the
main features that we considered for the analysis is the \emph{minimum frequency
of contact}. It defines, for the alters included in each circle, the lower bound
of the frequencies of contacts of their social links. In other words, this value
indicates the minimum frequency of contact for an alter to be included in a
given circle. In the table, we report the average value of this measure as ``min
freq.'', calculated for all the ego networks in terms of number of contacts per
month. The minimum frequencies of contact of offline ego networks have been
taken according to the definition discussed in Section~\ref{sec:background}:
\emph{once a week} for the support clique, \emph{once a month} for the sympathy
group and \emph{once a year} for the active network while, for the affinity
group, the minimum frequency of contact has not been defined yet.

In the table, we also show the average size of the obtained circles for online
ego networks while, for offline networks, we report the values presented
in~\cite{Zhou2005}, that summarise the properties of a large number of offline
social networks obtained in diverse social environments. Despite the size of the
circles in Facebook and Twitter ego networks appear to be very close to each
other, it is worth to remind that they should not be compared directly. In fact,
as already explained in Section~\ref{sec:download}, the ego networks in the
Facebook data set contain just a sample of the social relationships of the egos.
This is because the crawling process may have not downloaded the considered
regional network completely and because all the contacts external
to this area have
been excluded. In absence of precise information, we assume that the crawled
data represent a uniform random sample of both nodes and links. On the contrary,
the sizes of the circles of Twitter ego networks are more reliable, since we
have at our disposal the entire outgoing communication log of each ego (given
the limit of $3,200$ tweets).

Rather than the size, a better feature to consider to compare the properties of
online and offline ego networks is the scaling factor between the circles
(``scal. fact.'' in the table), defined as the ratio between the size of two
hierarchically adjacent circles. This measure can provide insights about how the
circles in ego network are hierarchically arranged and is not affected by a
random sampling of the links. In fact, with random sampling, the size of all the
circles changes proportionally without affecting the scaling factors. 

\subsection{Comparing Online and Offline Ego Networks} \label{sec:osnstructure}

\begin{table*}[t] \caption{Offline/online ego networks mapping.}
  \label{tab:mapping} \renewcommand{\arraystretch}{1.2} \centering \footnotesize
  \begin{tabular}{cl|cccccccccc|} \cline{3-12} & &
    \multicolumn{2}{c}{\parbox[c][10mm]{15mm}{\centering\textbf{Super support
    clique}}} &
    \multicolumn{2}{c}{\centering\parbox{15mm}{\centering\textbf{Support
    clique}}} & \multicolumn{2}{c}{\parbox{15mm}{\centering\textbf{Sympathy
    group}}} & \multicolumn{2}{c}{\parbox{15mm}{\centering\textbf{Affinity
    group}}} & \multicolumn{2}{c|}{\parbox{15mm}{\centering\textbf{Active
    network}}} \\
    \hline
    \multicolumn{1}{|c|}{\multirow{3}{*}{\textbf{Facebook}}} & circle &
    \multicolumn{2}{c}{$-$} & \multicolumn{2}{c}{$C_1$} &
    \multicolumn{2}{c}{$C_2$} & \multicolumn{2}{c}{$C_3$} &
    \multicolumn{2}{c|}{$C_4$} \\
    \multicolumn{1}{|c|}{} & min freq. &
    \multicolumn{2}{c}{$-$} & \multicolumn{2}{c}{$5.09$} &
    \multicolumn{2}{c}{$1.95$} & \multicolumn{2}{c}{$0.67$} &
    \multicolumn{2}{c|}{$0.11$} \\
    \multicolumn{1}{|c|}{} &
    size\tablefootnote{Scaled size to match offline active network dimension.} &
    \multicolumn{2}{c}{$-$} & \multicolumn{2}{c}{$(4.70)$} &
    \multicolumn{2}{c}{$(15.31)$} & \multicolumn{2}{c}{$(44.77)$} &
    \multicolumn{2}{c|}{$(132.50)$} \\
    \hline
    \multicolumn{1}{|c|}{\multirow{3}{*}{\textbf{Twitter}}} & circle &
    \multicolumn{2}{c}{$C_1$} & \multicolumn{2}{c}{$C_2$} &
    \multicolumn{2}{c}{$C_3$} & \multicolumn{2}{c}{$C_4$} &
    \multicolumn{2}{c|}{$C_5$}\\
    \multicolumn{1}{|c|}{}& min freq. &
    \multicolumn{2}{c}{$20.55$} & \multicolumn{2}{c}{$8.91$} &
    \multicolumn{2}{c}{$3.98$} & \multicolumn{2}{c}{$1.36$} &
    \multicolumn{2}{c|}{$0.18$}\\
    \multicolumn{1}{|c|}{}& size &
    \multicolumn{2}{c}{$1.66$} & \multicolumn{2}{c}{$5.06$} &
    \multicolumn{2}{c}{$12.87$} & \multicolumn{2}{c}{$32.66$} &
    \multicolumn{2}{c|}{$97.47$}\\
    \hline
    \multicolumn{1}{|c|}{\multirow{3}{*}{\textbf{Offline}}} & circle &
    \multicolumn{2}{c}{$-$} & \multicolumn{2}{c}{$C_1$} &
    \multicolumn{2}{c}{$C_2$} & \multicolumn{2}{c}{$C_3$} &
    \multicolumn{2}{c|}{$C_4$} \\
    \multicolumn{1}{|c|}{}& min freq. &
    \multicolumn{2}{c}{$-$} & \multicolumn{2}{c}{$4.29$} &
    \multicolumn{2}{c}{$1.00$} & \multicolumn{2}{c}{$-$} &
    \multicolumn{2}{c|}{$0.08$} \\
    \multicolumn{1}{|c|}{}& size &
    \multicolumn{2}{c}{$-$} & \multicolumn{2}{c}{$4.6$} &
    \multicolumn{2}{c}{$14.3$} & \multicolumn{2}{c}{$42.6$} &
    \multicolumn{2}{c|}{$132.5$} \\
    \hline
  \end{tabular}
\end{table*}

Looking at the scaling factors in Table~\ref{tab:clusters}, we can see that
their values are very similar to each other and close to $3$, for both Facebook
and Twitter ego networks, and they are compatible with the results found
offline. A scaling factor of three has been found in several offline social
networks and it appears to be a fundamental property of human ego
networks~\cite{Zhou2005}. This result is another indication that Facebook and
Twitter ego networks show a hierarchical structure remarkably similar to that
found in offline environments.

Considering the average minimum frequency of contact of the circles, we can
note that there is a match between the circles of the two OSNs and those of
offline social networks. Specifically, as we report in Table~\ref{tab:mapping},
we find the same magnitude in the ``min freq.'' values of $C_1$ in Facebook,
$C_2$ in Twitter and $C_1$ in offline social networks, that therefore we map to
the concept of support clique. In the same way, $C_2$ in Facebook can be matched
to $C_3$ in Twitter and $C_2$ in offline environments (the sympathy group),
$C_3$ in Facebook matches $C_4$ in Twitter, and we hypothesise that the two
match $C_3$ offline (affinity group). $C_4$ in Facebook matches $C_5$ in Twitter
and $C_4$ offline (the active network). It is worth noting that Twitter shows
higher values of min. freq (nearly double) for all the circles compared to
Facebook and offline ego networks. This could be ascribed to the nature of the
platform, and to the measure of interaction that we used, which could be slightly
different than the one used in the other environments.

Last, we have compared the ego networks according to the sizes of their layers,
which is another important signature of offline ego networks. The match between
$C_2$-$C_5$ in Twitter and $C_1$-$C_4$ offline is further confirmed by a strong
similarity in their size, as reported in Table~\ref{tab:mapping}. In the case of
Facebook, a direct comparison is not possible, because of the unknowns in the
sampling process previously discussed. Nevertheless, we can obtain strong hints
about a significant match by re-scaling the Facebook sizes, as follows. Assuming
that $C_4$ in Facebook matches $C_4$ offline (which is suggested considering the
minimum frequency and the scaling factors), we have re-scaled the size of $C_4$
in Facebook to match the size of $C_4$ offline ($132.50$). The resulting ratio
has a value of $2.63$ that we have applied to the other Facebook layers. Note
that the value of $2.63$ is compatible with the reported subsampling of other
networks obtained using the same crawling agent~\cite{Wilson2012}. It is
interesting to note that, scaling the size of other Facebook circles ($C_1$,
$C_2$ and $C_3$) according to this ratio, they match very well the respective
sizes of the offline layers. 

Interestingly, in Twitter we have found that there is an additional circle
($C_1$) with a very high minimum frequency of contact that represents a
subcircle of the support clique. Since the sizes of $C_2$-$C_5$ in
Twitter show a
good match with those found offline, we can say that $C_1$ in Twitter, which we
call ``super support clique'', has a typical size of $1$ or $2$ people. This
additional circle has been already hypothesised in offline social networks, but
its existence remained unconfirmed hitherto, due to absence of big
enough data sets to reliably highlight this type of
relationships~\cite{dunbarprivate}.

Summarising, our results show that there is a remarkable similarity between ego
networks in OSNs (both Facebook and Twitter) and offline networks, in terms of
scaling factors, minimum interaction frequency and size of the layers. This
suggests that the use of OSNs does not affect the structural properties of ego
networks, that are instead controlled by the constrained nature of
the human brain.
In addition our results also highlight additional structural elements, i.e. the
``super support clique'' in Twitter. This is a very interesting result per se,
and also shows that OSNs can be used as an extremely useful tool to collect
large-scale data sets to characterise human social network
properties. The scale at
which data can be collected with OSNs permits to draw statistically relevant
conclusions, which is often much harder or cumbersome with more conventional
data collection campaigns (such as standard questionnaires). From a more
technological standpoint, our results could be useful for the creation of
advanced social platforms and efficient networking solutions for the Future
Internet. For example, differences in the properties of social contacts of the
user, arranged into the ego network circles, could be exploited to automatically
set privacy policies (e.g. giving more trust to close friends) or to facilitate
the management of social relationships giving specific tools for each circle.

\section{Analysis of Information Diffusion in Twitter Ego Networks}
\label{sec:diffusion_analysis}

In this section, we assess the extent to which
ego network structural properties in OSNs impact on information diffusion inside
the ego networks. To do so, we analyse ego networks in Twitter, for which we
have accurate information regarding the creation of tweets and retweets and we
can thus understand how information is propagated by users, and we seek for the
relation between direct interactions of egos with their alters and the quantity
of information that these egos retweet from each social link. This kind of
analysis is clearly not possible in the Facebook data set, in which no exact
information is known about the data exchanged amongst users.

Since the Twitter data set contains information about complete ego networks and
also about all the tweets circulating in these networks, we can perform a
detailed analysis of one-hop information diffusion. Clearly, we do not have
enough data for the analysis of complete information cascades in Twitter, but
this would be a natural extension of our work that we are currently
investigating using additional Twitter data. In addition, we did not consider
the textual content of tweets, to see to what extent structural
  properties alone explain information diffusion patterns, without delving into
analysis of the content of exchanged messages.

In this work, we analysed the same Twitter ego networks that we already
described, in terms of structural properties, in Section~\ref{sec:analysis}. The
key idea that we used to assess the impact of the structure of ego networks in
the diffusion of information is to count the number of tweets originally
generated by each alter that are retweeted by the considered ego, relating this
measure to the tie strength of their social link, calculated - as previously
done - as the frequency of direct communication between them. 

All the information
diffusion mechanisms induced by sources external to ego networks, such as, for
example, the trending topic page of Twitter, are not considered. This allows us
to study the information diffusion derived from the presence of a social
relationship, eliminating possible bias derived from external sources. Of
course, external sources play an important role in the diffusion of information,
but this is out of the scope of the present work and has been already
characterised in Twitter~\cite{Myers2012}.

Before performing the analysis, we normalised the data by the duration of each
social link between users and by the differences between ego networks due to the
characteristics of the egos. This ensures a homogeneous analysis, eliminating
differences between ego networks due to their different duration or the
different frequency of use of the users. To normalise the data, we used, as a
measure of tie strength, the percentage of frequency of replies sent by ego to
her alter with respect to the total frequency of replies of ego to all her
alters during her entire lifespan. This measure is expressed by the following
equation.

\begin{equation}
  \label{eq:norm_freq}
  \overline{frep}_{e,a} =
  \frac{\mathrm{link\;reply\;frequency}}{\mathrm{ego\;total\;reply\;frequency}}
  = \frac{rep_{e,a}}{l_{e,a}} * \frac{lt_{e}}{reptot_{e}}
\end{equation}
where
$rep_{e,a}$ is the number of replies sent from ego ($e$) to alter ($a$),
$l_{e,a}$ is the lifespan of the social link between $e$ and $a$, $reptot_{e}$
is the total number of replies sent by $e$ to her alters, and $lt_{e}$ is the
lifespan of the ego. As already done in Section~\ref{sec:analysis}, the
frequency of contact (replies) is estimated by dividing the number of recorded
replies sent from the ego to the the considered alter by the time elapsed
between the first mention or reply sent by ego to the alter and the time of the
download (link lifespan). The lifespan of the ego is the time elapsed between
the creation of her account and the time of the download. Note that
$\overline{frep}$ is different from the normalised contact frequency used in
Section~\ref{sec:analysis}, where each contact frequency was divided by the
maximum contact frequency of the respective ego network to obtain a value in
[0,1]. Here the purpose of the normalisation is different and the obtained
measure is more suited for capturing the differences between egos, although it
does not necessarily result in a value in [0,1].

To measure information diffusion, we used the frequency of retweets of egos
generated from tweets of their alters, divided by the total frequency of
retweets of the egos, as defined by the following equation.

\begin{equation}
  \label{eq:norm_ret_freq}
  \overline{fret}_{e,a} =
  \frac{\mathrm{link\;retweet\;frequency}}{\mathrm{ego\;total\;retweet\;frequency}}
  = \frac{ret_{e,a}}{lret_{e,a}} * \frac{lt_{e}}{rettot_{e}}
\end{equation}
where $ret_{e,a}$ is the number of retweets by ego ($e$) of tweets originally
generated by her alter ($a$), $lret_{e,a}$ is the maximum between $l_{e,a}$ and
the time elapsed between the first retweet done by $e$ to a tweet originally
created by $a$ and the time of the download, $rettot_{e}$ is the total number of
retweets done by $e$ of tweets originally generated by her alters, and $lt_{e}$
is the lifespan of the ego.

As a first analysis of the relation between tie strength and information diffusion,
we calculated the correlation between $\overline{frep}$ and $\overline{fret}$ for
all the relationships belonging to the active network (C5) of the ego networks in
our Twitter data set and we
  also fitted a linear function relating the two measures, defined as follows.

\begin{equation}
  \label{eq:regression_equation}
  \overline{fret} = \alpha + \beta * \overline{frep}
\end{equation}

For the analysis, we have considered all the relationships of the ego networks together.
  Nevertheless, we also performed the same analysis by considering each ego network separately,
  and then averaging the results for all the egos. The results obtained with the two
  techniques are very similar, although for the second case the significance is
  sometimes not sufficient for ego networks with a small number of relationships. For this reason,
  in the following we report only the results obtained by taking all the ego networks together.

\begin{table*}[t] \caption{Information diffusion properties of ego network rings
    in Twitter, where x and y are $\overline{frep}$ and $\overline{fret}$.}
    \label{tab:rings_diffusion}
    \footnotesize
    \centering
    \begin{tabular}{|c|c|c|c|c|c|c|c|c|c|} \hline &
      \multicolumn{3}{c|}{\textbf{all alters}} &
      \multicolumn{3}{c|}{\textbf{soc. rel alters}} &
      \multicolumn{3}{c|}{\textbf{other alters}}\\ \hline \textbf{Ring} &
      $r_{xy}$ & $\hat{\beta}$ & $\hat{\alpha}$ & $r_{xy}$ & $\hat{\beta}$ &
      $\hat{\alpha}$ & $r_{xy}$ & $\hat{\beta}$ & $\hat{\alpha}$\\ \hline $R_1$
      & 0.61 & 0.49 & 0.03 & 0.80 & 0.74 & 0.03 & 0.74 & 0.58 & -0.01 \\ $R_2$ &
      0.52 & 0.62 & 0.01 & 0.76 & 0.76 & 0.02 & 0.71 & 0.59 & 0.02 \\ $R_3$ &
      0.44 & 0.74 & 0.00 & 0.72 & 0.80 & 0.03 & 0.67 & 0.64 & 0.02 \\ $R_4$ &
      0.34 & 0.97 & 0.00 & 0.66 & 0.85 & 0.06 & 0.65 & 0.72 & 0.02 \\ $R_5$ &
      0.22 & 1.58 & 0.00 & 0.61 & 0.99 & 0.09 & 0.65 & 0.93 & 0.03 \\ \hline
      \textbf{Whole net (C5)} & 0.46 & 0.57 & 0.02 & 0.68 & 0.83 & 0.09 & 0.65 & 0.78 & 0.03 \\ \hline
      
    \end{tabular}
  \end{table*}

The correlation between $\overline{frep}$ and $\overline{fret}$ ($r_{xy}$) and the
  estimated parameters $\alpha$ and
  $\beta$ are reported in the last row of Table~\ref{tab:rings_diffusion}, under the
  column ``all alters''.

\begin{figure}[t]
  \centering
  \includegraphics[width=0.4\textwidth]{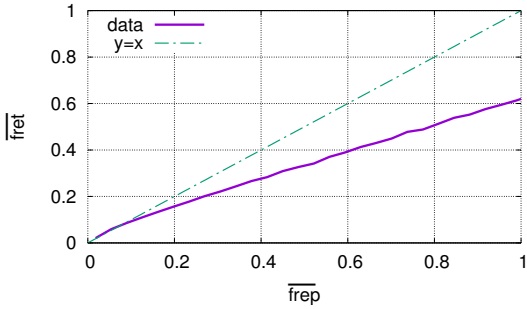}
  \caption{$\overline{fret}$ as a function of $\overline{frep}$.}
  \label{fig:xf_yf}
\end{figure}

The medium/high value of correlation ($r = 0.46$) is a first indication
of a relation between tie strength and information diffusion. To further analyse the impact of ego network structure on the relation between
$\overline{frep}$ and $\overline{fret}$, we
performed the same analysis by considering each layer of the ego networks separately.
To do so, we assigned each social link in the network to a position in the ego network model,
according to the contact frequency between the users it connects. Remember that,
by definition, the ego network model forms a hierarchical structure, and
therefore outer layers include inner ones. Thus, to avoid ambiguity, we have
assigned each link to a \emph{social ring}, defined as the part of a social
circle that is not included in any nested circles. To do so, we used the same
clustering technique described in Section~\ref{sec:clustanalysis}, considering
that the clusters coincide with social rings. A mapping between ego network
circles and rings is provided in Table~\ref{tab:rings}, where $R_1$ represents
the ego network ring containing alters with higher contact frequency, and $R_5$
is the outermost ring.

The results of the analysis for the different rings are reported in Table~\ref{tab:rings_diffusion}
  in the rows related to $R_1$-$R_5$. The correlation between $\overline{frep}$ and $\overline{fret}$ ranges
between $0.61$ for the innermost ego network ring and $0.22$ for the outermost
one (note that we are refferring here to the ``all alters'' column only). Compared to the average
correlation calculated on all the relationships
in the ego networks (equal to $0.46$), these values denote that the correlation
is higher for more internal rings, and decreases as we move from inner to outer rings.
This indicates that in the outer part
of the ego networks the two measures are less
dependent, and this could be explained by the fact that in this part of the
network alters are more heterogeneous. The lower correlation might also
  be due to the fact that for information coming from outer rings, the content
  of information is more important than the strength of the ties (also because
  in general tie strength is quite low on outer layers), and therefore the
retweeting behaviour is less correlated with social interactions.

The value of $\beta$ increases from inner to outer rings. In the inner
rings, a direct contact is related to less than one retweet ($\beta<1$), whilst in the
outer circles a direct contact is related to a higher number of
retweets ($\beta>1$). This is visible in Figure~\ref{fig:xf_yf}, which
depicts $\overline{fret}$ as a function of $\overline{frep}$.

\begin{table}[t]
  \caption{Ego network rings.} \label{tab:rings}
  \footnotesize
  \centering
  \begin{tabular}{|c|l|}
    \hline
    \textbf{Ring} & \textbf{Social
      circles correspondence\tablefootnote{Social circles are defined in
	Section~\ref{sec:microlevel}.}}\\ 
	\hline $R_1$ & super support clique\\
	$R_2$ & support clique, excluded the super support clique\\
	$R_3$ &
	sympathy group, excluded the support clique\\
	$R_4$ & affinity group,
	excluded the sympathy group\\
	$R_5$ & active network, excluded the
	affinity group\\
	\hline
      \end{tabular}
    \end{table}%

A possible explanation of the lower
  values of $\beta$ in the innermost layers could be that the relative gain in terms of
  information diffusion due to an increment in terms of tie strength may saturate after
  a certain level of strength (i.e., there is a sort of marginal utility
    law governing the dependency between tie strength and information diffusion).
  From the literature, we know that information coming
  from strong ties tends to remain trapped into highly clusterised parts of the network formed
  of nodes socially close to egos. Therefore, the information
  circulating between strong ties tend to be not very diverse, and thus an increment in
  terms of tie strength could not be accompanied by an equal increment of information diffusion.
  On the other hand, for lower values of tie strength, the information coming
  from alters
  is generally much more diverse, and this could explain a lower dependency
  upon tie strength, and thus higher values of $\beta$. In other
    words, the diversity of information coming from weak ties increases its
    probability of being retweeted, with respect to the sole effect of the
  strength of the social tie over which information arrives to the ego.
   Another possible explanation for this trend is that egos may choose their
   strong ties primarily on the basis of their
  emotional closeness, while weak ties could be picked primarily because
  of the information they allow the ego to access, which is therefore
proportionally more likely to be retweeted.

As described in Section~\ref{sec:datasets}, users in Twitter can be divided in
two distinct groups: socially relevant users and other users. We have already
seen the differences in terms of social behaviour of the users in these groups,
with the former containing people who use Twitter for socialising and
maintaining relationships with others, and the latter containing users with a
less ``human'' social behaviour, e.g. companies, public figures, bots, and
others. In the analysis performed in Section~\ref{sec:analysis}, we studied the
structure of the ego networks of socially relevant users, considering all their
alters, which could belong to both classes. Going deeply into detail into the
relation between information diffusion and the structural properties of ego
networks, we need to analyse the behavioural differences of socially relevant
users towards their different classes of alters. For this reason, the analysis
introduced in this section is performed not only on all alters without
distinction, but also considering socially relevant alters and other alters
separately. Unfortunately, we have complete information about the nature of an
alter only in case her profile has been downloaded by our crawler, so, for each
ego network, we only have a fraction of alters that we can classify, which is, on
average, $\sim30\%$ of the ego network. Nevertheless, the sample of classifiable
relationships for each ego network can be considered a random sample of the
relationships of egos. Therefore, the results presented for the different
classes are estimated by using the set of classified alters.
Moreover, as we consider all the social relationships of each
circle for all the ego networks in the data set, the number of
relationships is sufficient to obtain significant results.

In the Twitter ego networks that we have crawled we find that
there are, on average, $27.8\%$
socially relevant alters, and $72.2\%$ other alters. Note that, as already shown
in Section~\ref{sec:datasets}, the majority of egos in the initial data set are
socially relevant users. This means that, also according to the statistics
presented in Section~\ref{sec:datasets}, socially relevant users have less
connections than other users, and this results in a higher proportion of
connections towards ``other alters'' than to socially relevant alters, also for
socially relevant egos.

As shown in Table~\ref{tab:rings_diffusion}, the correlations
between $\overline{frep}$ and $\overline{fret}$ considering all the
  social relationships in the ego networks (C5)
  for the two classes of alters are
respectively $0.68$ for socially relevant alters and $0.65$ for other alters.
In addition, Table~\ref{tab:rings_diffusion} reports the statistics regarding the relation
between $\overline{frep}$ and $\overline{fret}$ for the different categories of
alters divided into the different social rings.

When considering socially relevant alters
and other alters separately, the correlation is significantly higher in both
cases then the case where
alters are taken altogether. This indicates that there are two separate processes underpinning the
relation between tie strength and information diffusion for the two classes,
and, when the processes are mixed together, this difference is less visible. The
different values of $\alpha$ and $\beta$ for the two classes support this hypothesis.
The higher and more homogeneous values of $\beta$ for socially
relevant users indicates that these alters are treated in a more homogeneous way
by egos across the different rings. The increasing value of $\beta$ for
both classes moving from internal to external rings, in addition, confirms the
same phenomena already discussed when analysing all alters at the same time.

\begin{figure}[t]
  \centering
  \includegraphics[width=0.4\textwidth]{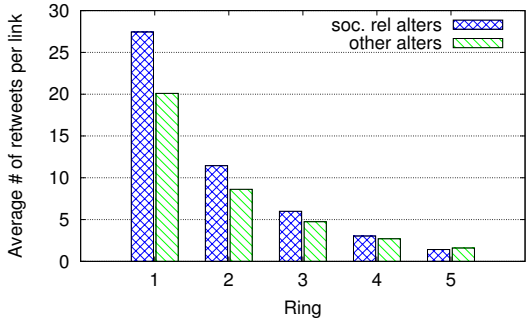}
  \caption{Average
  number of retweets per link divided by rings.}
  \label{fig:rings_diffusion}
\end{figure}

\begin{figure}[t]
  \centering
  \includegraphics[width=0.4\textwidth]{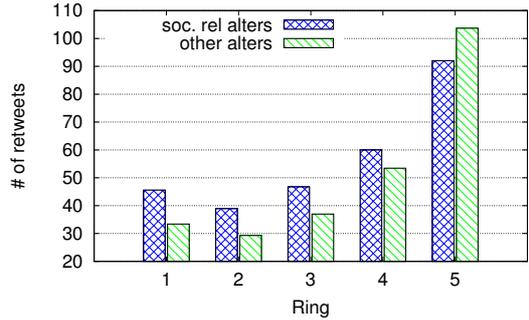}
  \caption{Average number of retweets per ego network, divided by ring.}
  \label{fig:rings_diffusion_weighted}
\end{figure}

Figure~\ref{fig:rings_diffusion} depicts the average number of retweets per
link, for the different rings. Inner circles show a higher number of retweets
per link, in accordance with the values of correlation and the estimated values
of the regressors of equation~\ref{eq:regression_equation}. It is worth noting
that this value, when multiplied by the average number of alters in each ring
(calculated from the size of the ego network circles obtained in
Section~\ref{sec:analysis}), indicates that, cumulatively, the quantity of
information diffused through the outer layers is higher than that in the inner
layers. This is visible in
Figure~\ref{fig:rings_diffusion_weighted}
and it is in accordance with the idea of ``the strength of weak
ties''~\cite{Granovetter2007}, as also empirically found in
Facebook~\cite{Bakshy2012}. Interestingly, the first three rings show
  approximately the same amount of
diffusion, whilst the outermost rings, R4 and R5, bring significantly higher
levels of diffusion. When we divide socially relevant alters and other types of
alters, we find that the amount of information diffused in the first four rings
coming from the former class is higher than that coming from the latter. In the
outermost ring, we found the opposite behaviour, with socially relevant alters
providing less information than other types of alters. This is perhaps not too
surprising, as we expect to find alters that are not socially relevant
prevalently in the outermost ring, where the level of emotional closeness or
intimacy with the alters is lower than more internal rings.

Finally, to understand also the relation between the properties of the
  individual egos and their information diffusion patterns, we analysed the relation between the activity of the
egos, defined as the sum of contact frequencies of the links of each ego network
(see Section~\ref{sec:analysis}), and the properties of tweets originally
generated by egos and of her retweets. The correlation between the activity of
the egos and the number of tweets they generate is $0.38$. By applying a
logarithmic transformation to both the activity and the number of tweets
generated by the egos we obtain a higher correlation ($0.51$), indicating a
non-linear relation between the two measures. The correlation between the
activity and the number of retweets generated by egos (after the logarithmic
transformation) is $0.38$. We also analysed the relation between activity of the
egos and the average popularity of their tweets, calculated as the number of
retweets they received. Activity on Twitter appears to be uncorrelated to
popularity, showing a correlation value of $0.004$.

\section{Conclusion} \label{sec:conclusions}

In this paper, we presented an
analysis aimed at characterising the micro-level properties of OSNs and to
understand how these properties influence the formation of macro-level social
phenomena, specifically the diffusion of information in the network following
the word-of-mouth effect.

As far as the structure of OSNs is concerned, we have found that online ego
networks show properties that are remarkably similar to those found in offline
social networks. Specifically, we have analysed two data sets containing
interaction data collected from Facebook and Twitter, that have been processed
to obtain the online ego networks of a large number of users. The results of our
analysis indicate that the structures of offline and online ego networks are
compatible. In fact, we have found that the typical number of social circles in
online ego networks is equal to $4$ and the scaling factor between
hierarchically adjacent circles is very close to $3$. Moreover, the characteristic
frequency of communication inside the circles is
comparable with that measured offline, and that the sizes of the circles are
very similar. These results are in line with the fundamental properties of human
social networks found offline.

Looking in detail at the properties of the circles obtained from Facebook and
Twitter, we matched them with those defined in sociology and anthropology. The
results indicate that the four circles in Facebook are directly
mapped with their offline equivalents. The higher richness of the Twitter data
set allowed us to discover an additional circle nested in the support clique,
that we called ``super support clique''. This circle is characterised by a very
high frequency of contact ($17.28$ interaction per month) and small size, with
one or two members on average. Alters inside this circle could be a partner
and/or a best friend of the ego. For a long time this result has been
hypothesised in sociology and psychology, but, for the lack of data, the
presence of this additional circle has never been experimentally demonstrated
until now.

Building on the results on ego network structures in OSNs, we performed an
information diffusion analysis assessing the impact of the different ego network
rings (i.e. portion of each circle not containing the other nested circles) on
the process. Specifically, we performed a correlation analysis on Twitter data
to assess the relation between direct contact frequency (of Twitter replies) and
the frequency of retweets passing through social links. The results indicate
that the two measures are highly correlated, with links in the internal ego
network layers showing the highest correlation. As a further refinement of the
analysis, we classified the alters of each ego network into ``socially relevant
users'' and ``other users'', and we calculated the correlations for these
classes separately. Interestingly, the correlations of both classes are higher
when taken separately rather than analysing them together. This could indicate
the presence of two separate processes governing the diffusion of information
for the two classes of alters. The different values of angular coefficients of
the function explaining the relation between the measures of tie strength and
information diffusion for the two classes, estimated through linear regression,
support the presence of these two separate processes. The correlations found for
socially relevant alters are high (higher than 0.8 for the innermost layers),
indicating that the diffusion of information can be accurately explained as a
function of tie strength.

A possible practical application of our results may be to use them to improve
existing information diffusion models. The knowledge about the role of ego
network rings in the diffusion process may lead to more representative synthetic
diffusion traces than traditional models. Differences in the structural
properties of ego networks (e.g. size, number of layers, tie strength
distribution) could also be useful for identifying influential information
spreaders in the network. Another possible application field for the results of
our analysis is related to distributed online social networks, an alternative to
OSNs based on peer-to-peer communications. DOSN users would probably like to
replicate their data on nodes that they trust and help
disseminate content coming primarily from these nodes~\cite{Arnaboldi2014TheRole,Arnaboldi2015Information}.
From the analysis presented in this paper,
we know that there is a significant influence of tie strength on
information propagation, and DOSN system could exploit knowledge about tie
strength between users to estimate the level of information diffusion, and
replicate it accordingly (an initial effort in this sense is presented
in~\cite{Conti2014Epidemic}.

\appendix

\section*{APPENDIX}

\setcounter{section}{0}

\section{Facebook and Twitter} \label{app:platforms}

In this section, we present a brief discussion
about the main features of Facebook and Twitter with particular regard to the
mechanisms they provide to the users to communicate with each other.

\subsection{Facebook}

Facebook is the most used online social networking service in the world, with
roughly $1.26$ billion users as of 2013. Facebook was founded in 2004 and is
open to everyone over 13 years old. Facebook provides several features to the
users. First, each user has a \emph{profile} which reports her personal
information and it is accessible by other users according to their permissions
and the privacy settings of the user. Connected to her profile, the user has a
special message board called \emph{wall}, which reports all the asynchronous
messages made by the user (\emph{status updates}) or messages received from
other users (\emph{posts}). Posts (that include status updates) can contain
multimedia information such as pictures, URLs and videos. Users can
\emph{comment} posts to create discussions around them. Comments have the same
format as posts. To be able to access the personal page of other users, a user
must obtain their \emph{friendship}. A friendship is a bi-directional relation
between two users. Once a friendship is established, the involved users can
communicate with each other and view their personal information - depending on
their privacy settings. The users can visualise the activity of their
\emph{friends} by using a special page called \emph{news feed}.

\subsection{Twitter}

Twitter is an online social networking and microblogging service founded in
2006, with more than $500$ million registered users as of
2012\footnote{According to Twitter CEO Dick Costolo in October 2012.}. In
Twitter, users can post short public messages (with at most $140$ characters)
called \emph{tweets}. All the users' tweets are accessible by other users,
unless the users' profiles are private or the access is restricted by other
specific settings. Users can also automatically receive notifications of new
tweets created by other users by ``following'' them (i.e. creating a
subscription to their notifications). People following a specific user are
called her \emph{followers}, whilst the set of people followed by the user are
her \emph{friends}.

Tweets can be enriched with multimedia content (i.e. URLs, videos, pictures) and
by using special text characters to insert additional information. Specifically,
a tweet can reference one or more users with a special mark called
\emph{mention}. Users mentioned in a tweet automatically receive a notification,
even though they are not followers of the tweet's author. Users can also
\emph{reply} to tweets. In this case, a tweet is generated with an implicit
mention to the author of the replied tweet. This implies that replies represent
directional communications. Replies often require additional effort in terms of
cognitive resources compared to other tweets since they presuppose that the user
creating the reply has read the tweet she is replying. Twitter has also a
private messaging system, however, since private messages are not publicly
accessible, we did not collected them in our data set. 

In addition to mentions and replies, Twitter provides a series of mechanisms for
broadcast communication that represent the most popular features of the
platform. First, all the tweets are automatically sent towards all the
followers of their authors. Moreover, tweets can also be \emph{retweeted}. A
user can make a retweet to forward a tweet it to all her followers. Each tweet
can be assigned to a topic through the use of a special character called hashtag
(i.e. ``\#'') placed before the text indicating the topic. Hashtags are used by
Twitter to classify the tweets and to obtain \emph{trending topics}.

\section{Classifier for the selection of socially relevant users in Twitter}
\label{app:classifier}

To build the supervised learning classifier used to
select socially relevant users from Twitter data set (see
Section~\ref{sec:datasets} for more details), we manually classified a sample of
$500$ accounts, randomly drawn from the data set, and we used this
classifications to train a Support Vector Machine~\cite{Vapnik1995SUpport}. This
SVM uses a set of $115$ variables: $15$ of them related to the user's profile
(e.g., number of tweets, number of following and followers, account lifespan)
and $100$ obtained from her timeline (e.g., percentage of mentions, replies and
retweets, average tweets length, number of tweets made using external
applications).

To test the generality of the SVM (i.e., the ability to categorise correctly new
examples that differ from those used for training), we took $10$ random
sub-samples of the training set, each of which contains $80\%$ of the entries,
keeping the remaining $20\%$ for testing. Then, we applied the same methodology
used to create the SVM generated from the entire training set on the $10$
sub-samples. Doing so, we obtained different SVMs, trained using different
sub-samples of the training set, and of which we were able to assess the
accuracy. The average accuracy of these SVMs can be seen as an estimate of the
accuracy of the SVM derived from the complete training set. Specifically, we
calculated the \emph{accuracy} index, defined as the rate of correct
classifications, and the \emph{false positives rate}, where false positives are
accounts wrongly assigned to the ``socially relevant user'' class. In our
analysis, we considered only users falling in the ``socially relevant users''
class, thus it is particularly important to minimise the false positive
rate\footnote{False negatives are ``socially relevant users'' with behaviour
similar to the subjects in the ``other users'' class. For this reason we
consider them as outliers, since our analysis is focused on Twitter average
users.}. Minimising the false negative rate is also important but less critical,
as false negatives result in a reduction of the number of users on which we base
our analysis.

The average accuracy of our classification system is equal to $0.813~[\pm0.024]$
and the average false positives rate is $0.083~[\pm0.012]$ (values between
brackets are $95\%$ confidence interval). These results indicate that we were
able to identify socially relevant people in Twitter with sufficient accuracy,
even if people have different behaviours and characteristics (e.g., different
culture, religion, age). Moreover, the false positive rate is quite low (below
$10\%$). The results are of the same magnitude as those found in a similar
classification performed in Twitter~\cite{Chu2010Who}.

\section{Frequency of contact estimation in Facebook} \label{app:facebook}

In
this section, we provide details about the procedure that we used to estimate the
frequency of contact between users in the Facebook data set described in
Section~\ref{sec:datasets}. As described in the text, the data set is divided
into snapshots representing four temporal windows containing the number of
interactions occurred between the users during the considered time period.

\subsection{Definitions} \label{sec:defin}

\begin{figure}[t]
  \centering
  \includegraphics[width=0.4\textwidth]{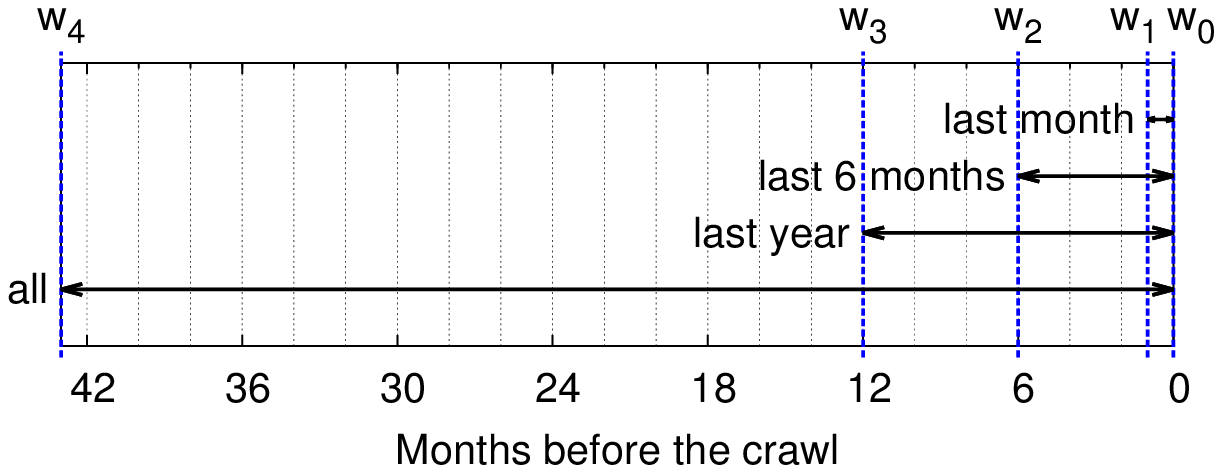}
  \caption{Temporal windows.}
  \label{fig:tempwind}
\end{figure}

We define the temporal window ``last month'' as the interval of time $(w_1,
w_0)$, where $w_1 = 1 \mbox{ month}$ (before the crawl) and $w_0 = 0$ is the
time of the crawl. Similarly, we define the temporal windows ``last six months'',
``last year'' and ``all'' as the intervals $(w_2, w_0)$, $(w_3, w_0)$ and $(w_4,
w_0)$ respectively, where $w_2 = 6 \mbox{ months}$, $w_3 = 12 \mbox{ months}$
and $w_4 = 43 \mbox{ months}$. $w_4$ is the maximum possible duration of a
social link in the data set, obtained by the difference between the time of the
crawl (April 2008) and the time Facebook started (September 2004). The different
temporal windows are depicted in Fig.~\ref{fig:tempwind}.

For a social relationship $r$, let $n_k(r)$ with $k \in \{1,2,3,4\}$ be the
number of interactions occurred in the temporal window $(w_k, w_0)$. Since all
the temporal windows in the data set are nested, $n_1 \leq n_2 \leq n_3 \leq
n_4$. If no interactions occurred during a temporal window $(w_k, w_0)$, then
$n_k(r) = 0$. As a consequence of our definition of active relationship, since
$n_4(r)$ refers to the temporal window ``all'', $n_4(r) > 0$ only if $r$ is an
active relationship, otherwise, if $r$ is inactive, $n_4(r) = 0$.

The first broad estimation that we can do to discover the duration of social ties in
the data set is to divide the relationships into different classes $C_k$, each
of which indicates in which interval of time $(w_k, w_{k - 1})$ the
relationships contained in it has started (i.e. the first interaction has occurred).
We can perform this classification by analysing, for each relationship, the number of
interactions in the different temporal windows. If all the temporal windows
contain the same number of interactions, the relationship must be born less than
one month before the time of the crawl, that is to say in the time interval
$(w_1, w_0)$. These relationships belong to the class $C_1$. Similarly,
considering the smallest temporal window (in terms of temporal size) that
contains the total number of interactions (equal to $n_4$), we were able to
identify social links with duration between one month and six months (class
$C_2$), six months and one year (class $C_3$), and greater than one year (class
$C_4$). The classes of social relationships are summarised in
Table~\ref{table:classes}.

\begin{table}[b] \renewcommand{\arraystretch}{1.2}
  \caption{Facebook classes of
  relationships.} \label{table:classes}
  \centering
  \footnotesize
  \begin{tabular}{|c|c|c|}
    \hline
    \textbf{Class} & \textbf{Time interval (in
    months)} & \textbf{Condition}\\
    \hline $C_1$ & $(w_1 = 1, w_0 = 0)$ & $n_1 =
    n_2 = n_3 = n_4$\\
    $C_2$ & $(w_2 = 6, w_1 = 1)$ & $n_1 < n_2 = n_3 = n_4$\\
    $C_3$ & $(w_3 = 12, w_2 = 6)$ & $n_1 \leq n_2 < n_3 = n_4$\\
    $C_4$ & $(w_4 =
    43, w_3 = 12)$ & $n_1 \leq n_2 \leq n_3 < n_4$\\
    \hline
  \end{tabular}
\end{table}

\subsection{Estimation of the Duration of the Social Links} \label{sec:estdur}

\begin{figure}[!t]
  \centering
  \includegraphics[width=0.4\textwidth]{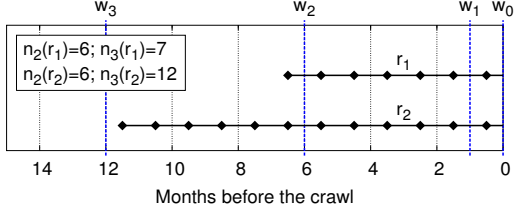}
  \caption{Graphical
  representation of two social relationships with different duration.}
  \label{fig:relexample}
\end{figure}

Although the classification given in the previous subsection is extremely useful
for our analysis, the uncertainty regarding the estimation of the exact moment
of the establishment of social relationships is still too high to obtain
significant results from the data set. For example, the duration of a social
relationship $r_3 \in C_3$ can be either a few days more than six months or a
few days less than one year. To overcome this limitation, for each relationship
$r$ in the classes $C_{k \in \{2,3,4\}}$, we estimate the time of the first
interaction comparing the number of interactions $n_k$, made within the smallest
temporal window in which the first interaction occurred $(w_k, w_0)$, with the
number of interactions ($n_{k - 1}$), made in the previous temporal window in
terms of temporal size $(w_{k-1}, w_0)$. If $n_{k}(r)$ is much greater than
$n_{k - 1}(r)$, a large number of interactions occurred within the time interval
$(w_k, w_{k - 1})$. Assuming that these interactions are distributed in time
with a frequency similar to that in the window $(w_{k - 1}, w_0)$, the first
occurred interaction must be near the beginning of the considered time interval.
On the other hand, a little difference between $n_{k}(r)$ and $n_{k - 1}(r)$
indicates that only few interactions occurred in the considered time interval
$(w_k, w_{k - 1})$. Thus, assuming an almost constant frequency of interaction,
the first contact between the involved users must be at the end of the time
interval. The example in Figure~\ref{fig:relexample} is a graphical
representation of this concept.

In the figure, we consider two different social relationships $r_1$, $r_2 \in
C_3$. The difference between the respective values of $n_2$ and $n_3$ is small
for $r_1$ and much larger for $r_2$. For this reason, fixing the frequency of
contact, the estimate of the time of the first interaction of $r_1$ is near to
$w_2$, while the estimate for $r_2$ results closer to $w_3$.

In order to represent the percentage change between the number of interactions
$n_k$ and $n_{k-1}$, we calculated, for each relationship $r \in C_k$, what we call
\textit{social interaction ratio} $h(r)$, defined as:

\begin{equation}
  h(r) = \left\{ \begin{array}{ll} n_k(r) / n_{k - 1}(r) -1 &
    \mbox{if } r \in C_{k \in \{2,3,4\}}\\ 1 & \mbox{if } r \in C_1 \end{array}
  \right.  . 
\end{equation}

If $r \in C_1$ we set $h(r) = 1$ in order to be able to perform the remaining
part of the processing also for these relationships. The value assigned to
$h(r)$ with $r \in C_1$ is arbitrary and can be substituted by any value other
than zero without affecting the final result of the data processing. Considering
that $n_k(r)$ is greater than $n_{k-1}(r)$ by definition with $r \in
C_{k\in{2,3,4}}$, the value of $h(r)$ is always in the interval $(0,
\infty)$\footnote{In case $n_{k-1}(r) = 0$, we set $n_{k-1}(r) = 0.3$. This
constant is the expected number of interactions when the number of interactions,
within a temporal window, is lower than $1$.}.

Employing the social interaction ratio $h(r)$, we define the function
$\hat{d}(r)$ which, given a social relationship $r \in C_k$, estimates the point
in time at which the first interaction of $r$ occurred, within the time interval
$(w_k, w_{k - 1})$:

\begin{equation}
  \hat{d}(r)=w_{k-1} + (w_k - w_{k-1}) \cdot \frac{h(r)}{h(r) +
  a_k}\qquad r \in C_k, \label{func:durfunc}
\end{equation}
where $a_k$ is a constant, different for each class of relationship $C_k$.

Note that the value of $\hat{d}(r)$ is always in the interval $(w_{k - 1},
w_k)$. The greater $h(r)$ - which denotes a lot of interactions in the time
window $(w_k, w_{k - 1})$ - the closer $\hat{d}(r)$ is to $w_k$. The smaller
$h(r)$, the closer $\hat{d}(r)$ is to $w_{k-1}$. Moreover, the shape of $\hat{d}(r)$
and the value of $a_k$ are chosen relying on the results
about the Facebook growth rate, available in~\cite{Wilson2012}. Specifically,
the distribution of the estimated links duration, given by the function
$\hat{d}(r)$, should be as much similar as possible to the distribution of the
real links duration, which can be obtained analysing the growth trend of
Facebook over time. For this reason, we set the constants $a_k$ in order to
force the average link duration of each class of relationships to the value that
can be obtained by observing the Facebook growth rate.
In~\cite{Arnaboldi2012techrep} we provide a detailed description of this step of
our analysis.

\subsection{Estimation of the Frequency of Contact} \label{sec:estfreq}

After the estimation of social links duration, we were able to calculate the
frequency of contact $f(r)$ between the pair of individuals involved in each
social relationship $r$:

\begin{equation}
  f(r) = n_k(r) / \hat{d}(r) \qquad r \in C_k.
\end{equation}

Previous research work demonstrated that the pairwise user interaction decays
over time and it has its maximum right after link
establishment~\cite{Viswanath2009}. Therefore, if we assessed the intimacy level
of the social relationships with their contact frequencies, this would cause an
overestimation of the intimacy of the youngest relationships. In order to
overcome this problem, we multiplied the contact frequencies of the relationships
in the classes $C_1$ and $C_2$ by the scaling factors $m_1$ and $m_2$
respectively, which correct the bias introduced by the spike of frequency close
to the establishment of the link. Assuming that the relationships established
more than six months before the time of the crawl are stable, we set $m_1$ and
$m_2$ comparing the average contact frequency of each of the classes $C_1$ and
$C_2$, with that for the classes $C_3$ and $C_4$. The obtained values of the scaling
factors are: $m_1 = 0.18$, $m_2 = 0.82$. Setting $m_3 = 1$ and $m_4 = 1$, the scaled
frequencies of contact are defined as:

\begin{equation}
  \hat{f}(r) = f(r) \cdot m_k \qquad r \in C_k.
\end{equation}

% Acknowledgments
%\section*{Acknowledgements} This work has been partially founded by the EU
%Commission under the projects\dots

\bibliographystyle{elsarticle-num}
\bibliography{paper}

\end{document}